\documentclass[journal]{IEEEtran}

\usepackage{cite}
\usepackage{graphicx}
 	\graphicspath{{../figures/}}
 	\DeclareGraphicsExtensions{.eps}
\usepackage[cmex10]{amsmath}
\usepackage{amssymb}
\usepackage{amsthm}
\usepackage{mathrsfs}
\usepackage{algorithmicx}
\usepackage{algpseudocode}
\usepackage{algorithm}
\usepackage{array}
\newtheorem{theorem}{Theorem}

\newtheorem{assumption}{Assumption}
\newtheorem{proposition}{Proposition}
\newtheorem{remark}{Remark}

\begin{document}
%
% paper title
% Titles are generally capitalized except for words such as a, an, and, as,
% at, but, by, for, in, nor, of, on, or, the, to and up, which are usually
% not capitalized unless they are the first or last word of the title.
% Linebreaks \\ can be used within to get better formatting as desired.
% Do not put math or special symbols in the title.
\title{An Introduction to Twisted Particle Filters and Parameter Estimation in Non-linear State-space Models}
%
%
% author names and IEEE memberships
% note positions of commas and nonbreaking spaces ( ~ ) LaTeX will not break
% a structure at a ~ so this keeps an author's name from being broken across
% two lines.
% use \thanks{} to gain access to the first footnote area
% a separate \thanks must be used for each paragraph as LaTeX2e's \thanks
% was not built to handle multiple paragraphs
%
\author{Juha~Ala-Luhtala,
	Nick~Whiteley,
	Kari Heine
	and Robert Pich\'{e}% <-this % stops a space
\thanks{J. Ala-Luhtala is with the Department of Mathematics, Tampere University of Technology,
PO Box 553, 33101, Tampere, Finland, e-mail: juha.ala-luhtala@tut.fi.}% <-this % stops a space
\thanks{N. Whiteley is with the School of Mathematics, University of Bristol, University Walk, Bristol, BS8 1TW, UK, e-mail: nick.whiteley@bristol.ac.uk}
\thanks{K. Heine is with the Department of Statistical Science, University College London, Gower Street, London, WC1E 6BT, UK, e-mail: k.heine@ucl.ac.uk}
\thanks{R. Pich\'{e} is with the Department of Automation Science and Engineering, Tampere University of Technology, e-mail: robert.piche@tut.fi.}}% <-this % stops a space
% \thanks{Manuscript received April 19, 2005; revised September 17, 2014.}

% make the title area
\maketitle

% As a general rule, do not put math, special symbols or citations
% in the abstract or keywords.
\begin{abstract}
Twisted particle filters are a class of sequential Monte Carlo methods recently introduced by Whiteley and Lee \cite{whiteley2014twisted} to improve the efficiency of marginal likelihood estimation in state-space models. The purpose of this article is to extend the twisted particle filtering methodology, establish accessible theoretical results which convey its rationale, and provide a demonstration of its practical performance within particle Markov chain Monte Carlo for estimating static model parameters. We derive twisted particle filters that incorporate systematic or multinomial resampling and information from historical particle states, and a transparent proof which identifies the optimal algorithm for marginal likelihood estimation. We demonstrate how to approximate the optimal algorithm for nonlinear state-space models with Gaussian noise and we apply such approximations to two examples: a range and bearing tracking problem and an indoor positioning problem with Bluetooth signal strength measurements. We demonstrate improvements over standard algorithms in terms of variance of marginal likelihood estimates and Markov chain autocorrelation for given CPU time, and improved tracking performance using estimated parameters.
\end{abstract}

% Note that keywords are not normally used for peerreview papers.
\begin{IEEEkeywords}
Particle filter, sequential Monte Carlo, particle MCMC, Gaussian state-space model, parameter estimation.
\end{IEEEkeywords}

\section{Introduction}

State-space models are applied to a wide variety of signal processing problems, especially in positioning, tracking and navigation \cite{barshalom2004estimation, sarkka2013bayesian, ristic2004beyond}.
These models need to be calibrated by inferring unknown parameters from data.
There are a variety of approaches to this inference problem, such as maximum likelihood (ML) or maximum a posteriori (MAP) estimation using the Expectation Maximization algorithm or Laplace approximations, Gaussian filtering based approximations, and state augmentation techniques \cite{schon2011system, cappe2005inference, sarkka2013bayesian}. In this paper we consider a Bayesian approach, which has the advantage of allowing prior information about parameters to be imparted, and a variety of estimates and measures of uncertainty to be reported based on the posterior distribution. By using a Markov chain Monte Carlo (MCMC) algorithm, e.g. Metropolis-Hastings (M-H) (see \cite{casella1999montecarlo} for an introduction), one can in principle explore the entire posterior, but in practice the design of efficient MCMC algorithm can be a challenging task.

A direct application of M-H to a state-space model requires evaluation of the marginal likelihood of data, which is a high-dimensional, analytically intractable integral in many cases of interest. However, this issue can be circumvented through the application of pseudo-marginal MCMC methods \cite{beaumont2003estimation, andrieu2009pseudo}, which allow MCMC algorithms yielding samples from the desired posterior to be constructed if an unbiased estimator of the marginal likelihood is available. Particle filters \cite{gordon1993novel} (see \cite{ristic2004beyond, sarkka2013bayesian} for overviews in the context of tracking applications) provide such an estimator, and the resulting MCMC scheme is known as a particle Markov chain Monte Carlo (PMCMC) method \cite{andrieu2010particle}. Typically the most substantial contribution to the overall cost of a PMCMC algorithm arises from the need to run a particle filter at each iteration of the MCMC sampler, and the performance of the sampler is sensitive to the variability of the marginal likelihood estimate which the particle filter delivers \cite{andrieu2015convergence}. This motivates the development of particle filters which can provide reliable marginal likelihood estimates at a low computational cost.

In this paper we develop new ``twisted particle filtering'' methodology, building from ideas recently introduced by \cite{whiteley2014twisted}. Twisted particle filters are purposefully designed to provide more reliable approximations of marginal likelihoods than standard particle filters, while preserving the lack-of-bias property which permits their use within PMCMC.

Unlike traditional approaches to improving the efficiency of particle filters which modify the proposal distribution on a per-particle basis \cite{doucet2000sequential} or employ auxiliary weights for resampling \cite{pitt1999filtering}, twisted particle filters are derived by applying a form of re-weighting to the particle system as a whole, using a so-called ``twisting'' function. The role of the twisting function is to incorporate information from the observations, possibly future and past, into the mechanism by which particles are propagated over time. The ability to choose different twisting functions introduces a degree of freedom into the design of the particle algorithm, leading naturally to questions of optimality.   In concrete terms, if the twisting function is chosen well, the twisted particle filter can estimate the marginal likelihood with greater accuracy than a standard particle filter, in turn allowing more efficient estimation inference for static parameters in the state-space model.

The investigations of \cite{whiteley2014twisted} focussed mainly on theoretical analysis of twisted particle filters, studying their asymptotic properties in the regimes where the number of particles tends to infinity and where the length of the time horizon grows, under probabilistic assumptions on the observation sequence and strong regularity conditions on the statistical model.

The objectives of this paper are to present new twisted particle filtering methodology, validate it theoretically, and demonstrate its application and effectiveness within PMCMC for inferring the parameters of state-space models. Our main contributions are as follows.

\subsubsection{Algorithms} We introduce a general formulation of twisted particle filters.  The first novel aspect of this formulation beyond that given in \cite{whiteley2014twisted}, is that it allows various resampling methods to be incorporated in twisted particle filters. In particular, we derive a twisted particle filter around the popular systematic resampling method, which is known to reduce variance within the particle filter. The second novel aspect of the new formulation is that it allows for twisting functions which depend on historical particle states, which is an important factor when designing them efficiently in practice. The methodology of \cite{whiteley2014twisted}, which treated only multinomial resampling and twisting functions which depend only on current particle states. The utility of these algorithmic developments is that they allow for more accurate estimation of marginal likelihoods.

\subsubsection{Theory} We provide novel theoretical results which justify the new algorithms and characterize their optimal operation. The first result, Theorem \ref{thm:twisted_unbiased}, establishes the lack-of-bias property of the marginal likelihood approximation delivered by the new twisted particle filter. The importance of this result is that it justifies the use of the twisted particle filter within PMCMC, whilst allowing for more general structure of the resampling technique and twisting function than in \cite{whiteley2014twisted}.

The second result, Theorem \ref{thm:optimal}, identifies the twisting functions  which are optimal for approximating the marginal likelihood for a given finite number of observations. This provides a different perspective to the results of \cite{whiteley2014twisted}, which are asymptotic in nature, considering optimality in terms of minimal variance growth rate in the regime where the length of the data record tends to infinity.  Theorem \ref{thm:optimal} relies on only mild regularity assumptions on the ingredients of the twisted particle filter, whereas the results of \cite{whiteley2014twisted} assume a particularly strong form of geometric ergodicity of the signal in the hidden Markov model (HMM) and certain uniform upper and lower bounds on the likelihood functions. Moreover, compared to the analyses of \cite{whiteley2014twisted}, the proof of Theorem \ref{thm:optimal} is less intricate, and gives the reader a more accessible route to understanding how twisted particle filters work.

\subsubsection{Approximation techniques} Informed by Theorem 2, we propose methods to approximate the optimal twisting function for nonlinear Gaussian state-space models, based on ideas of Kalman filtering methodology together with local linearization using historical particle states.

\subsubsection{Applications and numerical results}  We provide numerical results in the context of two applications.

The first application is a range and bearing tracking problem. This is a classical nonlinear tracking scenario and serves as a benchmark application of particle filters \cite{gustafsson2002particle}, \cite{barshalom2004estimation}. The purpose of this example is to compare the performance of twisted particle filters and the corresponding PMCMC algorithms to standard particle filters in a situation where the ground truth for static parameters is available, with simulated data. The twisted particle filters we consider employ linearization techniques to approximate the optimal twisting functions.  The results we obtain illustrate that twisted particle filters can more reliably approximate marginal likelihoods for the same or less computational cost than standard particle filters. The benefits of using twisted particle filters within PMCMC are also demonstrated in terms of lower auto-correlation, and consequently more accurate approximation of posterior distributions over static parameters. We also compare tracking performance based on estimated parameter values.

The second application is a more complex indoor positioning problem. In this application a state-space model represents the unknown position of a user over time, observed indirectly and with uncertainty through received signal strength (RSS) measurements. Such data are widely available from many different wireless communication systems including mobile networks and WLAN. They have been proposed for use in location estimation in a variety of novel location-aware applications, such as environmental and structure monitoring, and many military and public safety applications, see \cite{patwari2005locating}, \cite{xinrong2006rss} and references therein. We work with a real Bluetooth RSS data set. A key task when dealing with RSS measurements is to calibrate the model by estimating unknown parameters which describe attenuation characteristics of the environment in which the user moves, since otherwise one must resort to oversimplified models \cite{xinrong2006rss}, which exhibit inferior tracking performance.

A variety of approaches to estimating these parameters have been suggested, involving least squares \cite{xinrong2006rss} and weighted least squares \cite{wang2012on} methods. These techniques provide point estimates of parameter values from a batch of data. Bayesian approaches, e.g., \cite{seshadri2005bayesian}, allow additionally for uncertainty associated with estimates to be reported, and incorporate prior distributions to quantify expert knowledge and physical constraints on parameter values. They also naturally handle uncertainty over state variables when inferring parameters through marginalization.

The price to pay for the Bayesian approach is the computational cost of Monte Carlo sampling, and so our numerical investigations largely focus on computational efficiency. We compare the performance of twisted particle filters to more standard particle filters using a variety of proposal and resampling techniques. We demonstrate improved CPU-time efficiency in estimating marginal likelihoods, and we show that this efficiency is carried over to the particle MCMC algorithm, giving performance gains in terms of quality of the resulting MCMC chain compared to PMCMC using a standard particle filter. We also demonstrate favourable tracking performance using parameter estimates obtained from PMCMC.

The structure of the paper is as follows.
Section \ref{sec:problem_formulation} gives the problem formulation.
Section \ref{sec:pmcmc} introduces a PMCMC algorithm and a standard particle filter.
Section \ref{sec:twisted_pf} presents the twisted particle filtering methodology and Theorems \ref{thm:twisted_unbiased}-\ref{thm:optimal}, which characterize the lack-of-bias property and optimal twisting functions. Computational complexity is also discussed.
Section \ref{sec:twisted_for_Gaussian} introduces methods for approximating the optimal twisting functions in nonlinear state-space models with Gaussian noise.
Section \ref{sec:applications} contains applications and numerical results.
Finally, conclusions are presented in Section \ref{sec:conclusion}.

\section{Problem formulation}\label{sec:problem_formulation}
We first introduce some notation.
Uppercase is used to denote random variables (e.g. $X, Y, \ldots$) and realized values are denoted with lowercase (e.g. $x,y,\ldots$).
For any sequence $(a_n)_{n \geq 0}$ and $s\leq k$ we write $a_{s:k}:=(a_s,\ldots,a_k)$.

We consider state-space models of the form
\begin{align}
X_0 \sim \mu_{0, \theta}(\cdot), \quad   X_k   & \sim f_{k,\theta}( \cdot \, | \, X_{k-1}), \quad k\geq1,\nonumber \\
 Y_k  & \sim g_{k,\theta}(\cdot \, | \, X_k), \quad k \geq 0, \label{eq:model}
\end{align}
where $X_k \in \mathbb{X}$ is the state vector, $Y_k \in \mathbb{Y}_k$ is the measurement vector, $\mu_{0,\theta}(\cdot)$ is the initial distribution, $f_{k,\theta}(\cdot \, | \, x_{k-1})$ describes the transitions of the state process and $g_{k,\theta}(\cdot \, | \, x_k)$ is the conditional distribution for the measurement.
All the model distributions are assumed to admit probability densities denoted with the same letter as the distribution.
The joint density of the state-variables and measurements for $k \geq 1$ is given by
\begin{align}
p_{\theta}(y_{0:k}, x_{0:k})& =\mu_{0,\theta}(x_0) g_{0,\theta}(y_0 \, | \, x_0) \nonumber \\
 & \quad \cdot\prod_{s=1}^k f_{s, \theta}(x_s \, | \, x_{s-1}) g_{s,\theta}(y_s \, | \, x_s).
\end{align}
The parameter vector $\theta \in \mathbb{R}^{d_{\theta}}$ contains all the unknown parameters of the model.

We are mainly concerned in estimating the unknown parameters $\theta$ using a set of realized measurements $y_{0:t}$.
In the Bayesian framework, the parameters are considered as random variables and estimates are computed using the posterior distribution
\begin{equation}
p(\theta \,|\, y_{0:t}) \propto p_{\theta}(y_{0:t}) p(\theta),
\label{eq:theta_post}
\end{equation}
where $p_{\theta}(y_{0:t})$ is the likelihood and $p(\theta)$ is the prior.

With the shorthand
$$
\pi_{k,\theta}^{-}(dx_k):=p_{\theta}(dx_k\, | \, y_{0:k-1}),\;\; \pi_{k,\theta}(dx_k):=p_{\theta}(dx_k\, | \, y_{0:k}),
$$
the likelihood term can be evaluated recursively, for $k \geq 1$,
\begin{align}
p_{\theta}(y_{0:k}) & = p_{\theta}(y_{0:k-1})\int_{\mathbb{X}} g_{k,\theta}(y_k \,|\, x_k) \pi_{k,\theta}^{-}(dx_k),\label{eq:likelihood_recursion}
\end{align}
$p_{\theta}(y_0) = \int_{\mathbb{X}} g_{0,\theta}(y_0 \, | \, x_0) \mu_{0,\theta}(dx_0)$,
and
\begin{equation}
\pi_{k,\theta}^-(x_k ) = \int_{\mathbb{X}}  f_{k,\theta}(x_k \,|\, x_{k-1}) \pi_{k-1,\theta}(dx_{k-1}), \;k \geq 1, \label{eq:predictive}
\end{equation}
\begin{equation}
\pi_{k,\theta}(x_k ) \propto \left\{ \begin{array}{cc} g_{0,\theta}(y_0 \, | \, x_0)\mu_{0,\theta}(x_0), & k=0, \\
 g_{k,\theta}(y_k \, | \, x_k)\pi_{k,\theta}^-(x_k), & k \geq 1. \end{array} \right. \label{eq:filtering}
\end{equation}

Exact inference using  (\ref{eq:theta_post}) directly is usually intractable, since the likelihood term can be evaluated exactly for only some special models (e.g. linear Gaussian model).
We consider particle filtering methods for computing unbiased estimates for the likelihood term.
These can then be used as a part of particle MCMC methods that draw samples from the posterior distribution of interest.

\section{Particle MCMC}\label{sec:pmcmc}

In this section we describe methods for drawing samples from the parameter posterior distribution in  (\ref{eq:theta_post}).
Algorithms targeting only the parameter posterior are often called marginal algorithms, because samples are drawn only from the marginal posterior $p(\theta \, | \, y_{0:t})$ instead of the full posterior $p(x_{0:t}, \theta \, | \, y_{0:t})$.

MCMC methods generate samples from the target posterior distribution by simulating a Markov chain $\theta^0,\theta^1,\ldots$ that has the target posterior distribution as a stationary distribution \cite{casella1999montecarlo}.
One of the best known and general MCMC methods is the Metropolis-Hastings (MH) algorithm, where a new sample $\theta^*$ at step $i$ is generated from a proposal distribution $\kappa(\cdot \, | \, \theta^{i-1})$.
The generated sample $\theta^*$ is then accepted with probability
\begin{equation}
\min \left\{ 1, \frac{p_{\theta^*}(y_{0:t})p(\theta^*)}{p_{\theta^{i-1}}(y_{0:t}) p(\theta^{i-1})}\frac{\kappa(\theta^{i-1} \,|\, \theta^*)}{\kappa(\theta^* \,|\, \theta^{i-1})}\right\}.
\end{equation}

To compute this acceptance probability, we need to evaluate likelihood terms $p_{\theta}(y_{0:t})$, but that is not possible for a general nonlinear state-space model. However, if an unbiased estimator for the likelihood is available, it is still possible to construct an MCMC algorithm to sample from the posterior distribution \cite{beaumont2003estimation, andrieu2009pseudo}.
For state-space models, we can use particle filters as unbiased estimators of the likelihood \cite{andrieu2010particle}.
A Metropolis-Hastings algorithm using particle filters to estimate the likelihood terms, called particle marginal Metropolis-Hastings (PMMH) \cite{andrieu2010particle}, is given in Algorithm \ref{alg:pmmh}.

\renewcommand{\figurename}{\textbf{Algorithm}}
\begin{figure}[!t]
\begin{algorithmic}[1]
\State Sample $\theta^{0} \sim p(\theta)$
\State Obtain an unbiased estimate $Z^{0}$ of $p_{\theta^{0}}(y_{0:t})$
\For{$i\geq 1$}
\State Sample $\theta^* \sim \kappa(\cdot \,|\, \theta^{i-1})$
\State Obtain an unbiased estimate $Z^*$ of $p_{\theta^*}(y_{0:t})$
\State Set $\alpha = \min \left\{1, \dfrac{Z^{*}p(\theta^*)}{Z^{i-1} p(\theta^{i-1})}\dfrac{\kappa(\theta^{i-1} \,|\, \theta^*)}{\kappa(\theta^* \,|\, \theta^{i-1})} \right\}$
\State Sample $U$ from a uniform distribution on $[0,1]$
\If{$U < \alpha$}
\State Set $\theta^{i} = \theta^*$ and $Z^{i} = Z^{*}$
\Else
\State Set $\theta^{i} = \theta^{i-1}$ and $Z^{i} = Z^{i-1}$
\EndIf
\EndFor
\end{algorithmic}
\caption{Particle marginal Metropolis-Hastings}
\label{alg:pmmh}
\end{figure}

\subsection{Particle filtering}

We proceed with an account of a standard particle filter. Our notation is in some places a little non-standard, but is chosen deliberately to help with the derivation of twisted particle filters in Section \ref{sec:twisted_pf}. Henceforth, for notational simplicity, we often omit the subscript $\theta$ and implicitly assume that the distributions can depend on the parameters.

We denote the set of $n\geq1$ particles at time $k\geq0$ by $\xi_k=(\xi_k^i)_{i=1}^n$, with corresponding unnormalized weights $W_k=(W_k^i)_{i=1}^n$. The filtering distribution is approximated by
\begin{equation}
\pi_{k,\theta}(d x_k) \approx \frac{\sum_{i=1}^n W_k^i \delta_{\xi_k^i}(d x_k)}{\sum_{i=1}^n W_k^i},
\end{equation}
where $\delta_{\xi_k^i}(\cdot)$ denotes a unit point mass centered at $\xi_k^i$.

In order to describe the sampling mechanism for the particles and understand certain properties of the algorithm it is convenient to also introduce,
for each $k\geq0$, the ancestor indicator variables $A_k=(A_k^i)_{i=1}^n$, where each $A_k^i$ takes a value in $\{1,\ldots,n\}$. If we also define for each $k\geq0$ and $i\in\{1,\ldots,n\}$, $(B_{k,j}^i)_{j=0}^k$ by letting $B_{k,k}^i:=i$ and for $k>0$,
recursively $B_{k,j}:=A_j^{B_{k,j+1}}$, $j=k-1,\ldots,0$, then we can write the ``ancestral line'' of particle $\xi_k^i$ as
\begin{equation}
\mathscr{L}_k^i:=(\xi_k^i,\xi_{k-1}^{B_{k,k-1}^i},\ldots,\xi_0^{B_{k,0}^i}),\label{eq:L_defn}
\end{equation}
which is a $\mathbb{X}^{k+1}$-valued random variable.

\begin{figure}[!t]
\begin{algorithmic}[1]
\For{$1 \leq i \leq n$}
\State Sample $\xi_0^{i} \sim q_0(\cdot)$
\State Set $W^i_0 = g_0(y_0 \, | \, \xi_0^i)\mu_0(\xi_0^i)/q_0(\xi_0^i)$
\EndFor
\State Set $Z_0 = \frac{1}{n}\sum_{i=1}^n W_{0}^i$
\For{$1 \leq k \leq t$}
\State Sample $U_{k-1} \sim \mathcal{U}[0,1]^m$
\State Set $A_{k-1}= r(U_{k-1},W_{k-1})$
\For{$ 1 \leq i \leq n$}
\State Sample $\xi_k^i \sim q_k(\cdot\, | \, \mathscr{L}_{k-1}^{A_{k-1}^i})$
\State Set $W_k^i = \dfrac{g_k(y_{k}\, | \, \xi_{k}^i) f_{k}(\xi_{k}^i \, | \, \xi_{k-1}^{A_{k-1}^i})}{q_k(\xi_k^i \, | \, \mathscr{L}_{k-1}^{A_{k-1}^i})}$
\EndFor
\State Set $Z_k = Z_{k-1} \frac{1}{n}\sum_{i=1}^n W_{k}^i$
\EndFor
\end{algorithmic}
\caption{Particle filter}
\label{alg:particle_filter}
\end{figure}
A particle filter is given in Algorithm \ref{alg:particle_filter}. Here
the proposal distributions $(q_k)_{k\geq0}$ are assumed to be chosen such that for each $k\geq0$ the weights $W_k$ are strictly positive and finite.
Each $q_k$ may be chosen to depend also on the observations $y_{0:k}$, but this dependence is suppressed from the notation.

\subsection{Resampling}

Lines 7 and 8 in Algorithm \ref{alg:particle_filter} together implement a generic resampling operation. Line 7 generates $U_{k-1}=(U_{k-1}^i)_{i=1}^m$ consisting of $m\geq1$ i.i.d. random variables, each uniformly distributed on $[0,1]$. Line 8 passes $U_{k-1}$ and the unnormalized weights $W_{k-1}$ to a deterministic mapping  $r \colon [0,1]^m \times \mathbb{R}_+^n \to \{1,\ldots,n\}^n$, which returns the ancestor indicator variables $A_{k-1}=(A_{k-1}^i)_{i=1}^n$. With $r^i(U_{k-1}, W_{k-1}) $ indicating the $i$th element in the vector returned by $r$, for brevity we sometimes write $r_{k-1}^i(U_{k-1}) \equiv  r^i(U_{k-1}, W_{k-1})$.

A variety of resampling mechanisms can be cast in this form through specific choices of $m$ and $r$.
We describe here two well known schemes: the multinomial and systematic methods; see \cite{douc2005comparison}
for background information. These techniques are standard; the details are included here in order to prepare for the presentation of the non-standard resampling techniques in twisted particle filters.

\begin{enumerate}
\item \textbf{Multinomial resampling}:
We have $m=n$ and the mapping $r$ is defined as
\begin{equation}
r^i(u, w) = j \quad \Leftrightarrow \quad u^i \in (d^{j-1}, d^j],
\label{eq:multinomial}
\end{equation}
where $d^0 = 0$ and $d^i = \sum_{j=1}^i w^j /(\sum_{j=1}^n w^j)$.

\item \textbf{Systematic resampling}:
We have $m=1$ and the mapping $r$ is defined as
\begin{equation}
r^i(u, w) = j \quad \Leftrightarrow \quad u+i-1 \in (nd^{j-1}, nd^j],
\label{eq:systematic}
\end{equation}
where $d^0 = 0$ and $d^i = \sum_{j=1}^i w^j /(\sum_{j=1}^n w^j)$.
\end{enumerate}

Systematic resampling is computationally light and has been found to have good empirical performance, although theoretical analysis is difficult due to high dependence between the resampled particles. Nevertheless, it is known, see e.g.  \cite{douc2005comparison}, that both multinomial and systematic resampling satisfy Assumption \ref{ass:resampling} below.

We define the shorthand notation $\mathscr{F}_0:=\xi_0$ and for $k\geq1$, $\mathscr{F}_k:=(\xi_0,U_0,\xi_1,\ldots,U_{k-1},\xi_{k})$.
\begin{assumption}\label{ass:resampling}
The mapping $r$ is such that for any $k\geq0$ and integrable function $\varphi:\mathbb{X}^{k+1}\to\mathbb{R}$,
$$
\mathbb{E}\left[\left.\frac{1}{n}\sum_{i=1}^{n} \varphi(\mathscr{L}_k^{r_k^i(U_{k})})  \right| \mathscr{F}_k  \right] = \frac{\sum_{i=1}^n W_k^i \varphi(\mathscr{L}_k^i)}{\sum_{i=1}^n W_k^i},
$$
where $\mathbb{E}$ denotes expectation when sampling according to Algorithm \ref{alg:particle_filter}.
\end{assumption}

Lines 5 and 13 compute a sequence $(Z_k)_{k=0}^t$, where each $Z_k$ is an estimate of $p(y_{0:k})$.
The following proposition justifies the use of Algorithm \ref{alg:particle_filter} to provide an unbiased estimate of $p(y_{0:t})$ at line 5 of Algorithm \ref{alg:pmmh}. This kind of result is well known; a proof is outlined in Appendix \ref{app:proof_prop1} for completeness.
\begin{proposition}\label{prop:unbiased}
If Assumption \ref{ass:resampling} holds, then for each $k\geq0$, $\mathbb{E}[Z_k]=p(y_{0:k})$.
\end{proposition}

\section{Twisted particle filters}\label{sec:twisted_pf}

In order to introduce and validate twisted particle filters we think more explicitly about $\xi_0$ and the sequence $(\xi_k,U_{k-1})_{k\geq1}$  as a stochastic process and consider the following initial and conditional distributions, according to which $\xi_0$ and $(\xi_k,U_{k-1})_{k\geq1}$ evolve when sampled through Algorithm \ref{alg:particle_filter}.
\begin{subequations}
  \begin{align}
  &\mathbf{M}_0(d \xi_0)  =  \prod_{i=1}^n q_0(d \xi_0^i), \label{eq:particle_filter_0} \\
 & \mathbf{M}_k(d \xi_k, d{u_{k-1}} \, | \, \mathscr{F}_{k-1}) \nonumber \\
 & \quad \quad \quad \quad =\mathcal{U}(du_{k-1}) \prod_{i=1}^n q_k(d\xi_k^i \, | \, \mathscr{L}_{k-1}^{r_{k-1}^i(u_{k-1})}),
\label{eq:particle_filter_k}
\end{align}
\end{subequations}
where $\mathcal{U}(du)$ denotes the uniform distribution on $[0,1]^m$.

Twisted particle filters are obtained by sampling the process  $\xi_0$, $(\xi_k,U_{k-1})_{k\geq1}$ from alternatives to \eqref{eq:particle_filter_0}--\eqref{eq:particle_filter_k}, which we discuss in more detail below.
\begin{remark}
For historical perspective, we note that the idea of constructing alternative distributions over the random variables in particle filters appears in some of the theoretical arguments which justify PMCMC \cite{andrieu2010particle}. However, the specifics of twisted particle filters are more akin to eigenfunction changes of measure for branching processes, which were studied earlier in the stochastic processes literature, see \cite[Section 3]{athreya2000change} and references therein.
\end{remark}

Let $(\psi_k)_{k\geq0}$ be a sequence of strictly positive functions, such that $\psi_0 \colon \mathbb{X}\to\mathbb{R}_+$ and for $k\geq1$, $\psi_k \colon \mathbb{X}^{k+1}\rightarrow\mathbb{R}_+$. We shall often write interchangeably $\psi_k(x_{0:k-1},x_k)\equiv\psi_k(x_{0:k})$. Each $\psi_k$ may also depend on $\mathscr{F}_{k-1}$ and any number of the measurements $y_k$, but this dependence is suppressed from the notation.

The initial and conditional distributions for the twisted particle filter are given by
\begin{subequations}
\begin{align}
& \widetilde{\mathbf{M}}_0(d \xi_0)  \propto \frac{1}{n} \sum_{s=1}^n \mathbf{M}_0(d \xi_0) \psi_0(\xi_0^s), \label{eq:particle_filter_t0}\\
& \widetilde{\mathbf{M}}_{k}(d \xi_k, du_{k-1} \, | \, \mathscr{F}_{k-1})  \nonumber  \\
& \quad \propto \frac{1}{n}\sum_{s=1}^n\mathbf{M}_k(d \xi_k, d{u_{k-1}} \, | \, \mathscr{F
}_{k-1})\psi_{k}(\mathscr{L}_{k-1}^{r_{k-1}^s(u_{k-1})},\xi_k^s), \label{eq:particle_filter_tk}
\end{align}
\end{subequations}
where the functions $\psi_k$ are called ``twisting functions''.
To avoid some tangential complications we shall assume henceforth that $\sup_x\psi_k(x)<\infty$ for each $k\geq0$, which is sufficient to ensure that the integrals needed to normalize $\widetilde{\mathbf{M}}_0$ and each $\widetilde{\mathbf{M}}_{k}$ are finite.

A more explicit expression for $\widetilde{\mathbf{M}}_0$  is obtained by plugging in \eqref{eq:particle_filter_0} and normalizing, to give
$$
\widetilde{\mathbf{M}}_0(d \xi_0) = \frac{1}{n}\sum_{s=1}^n \widetilde{q}_0(d\xi_0^s)\prod_{i\neq s}q_0(d\xi_0^i),
$$
where $\widetilde{q}_0(d\xi_0^s) := \psi_0(\xi_0)q_0(d\xi_0^s)/ \int\psi_0(x) q_0(dx)$. So to sample from $\widetilde{\mathbf{M}}_0$, one first draws a random variable, say $S_0$, from the uniform distribution on $\{1,\ldots,n\}$, then samples $\xi_0^{S_0}\sim \widetilde{q}_0(\cdot)$ and $\xi_0^{i}\sim q_0(\cdot)$ for $i\neq S_0$. Deriving a similar sampling recipe for $\widetilde{\mathbf{M}}_k$ is somewhat more involved. We state the resulting procedure in Algorithm \ref{alg:twisted_pf}, then formalize its validity and other properties in Theorems \ref{thm:twisted_unbiased} and \ref{thm:optimal}.

To write out Algorithm \ref{alg:twisted_pf} we need a few more definitions.
For $k \geq 0$, define the twisted (unnormalized) weights
\begin{equation}
\widetilde{W}_k^i:=W_k^i \widetilde{V}_k^i, \quad 1 \leq i \leq n, \label{eq:twisted_w}
\end{equation}
where
\begin{equation}
\widetilde{V}_k^i :=\int_{\mathbb{X}} \psi_{k+1} (\mathscr{L}_{k}^{i},x_{k+1} )q_{k+1}(dx_{k+1}\,|\,\mathscr{L}_{k}^{i}).
\label{eq:twisted_v}
\end{equation}
For $k\geq1$, define the twisted proposal distribution
\begin{equation}
\widetilde{q}_k(dx_k \, | \, x_{0:k-1}) \propto \psi_k(x_{0:k}) q_k(dx_k \, | \, x_{0:k-1}). \label{eq:twisted_q}
\end{equation}
Let $S_k$ be a discrete random variable conditional on $(\mathscr{L}_{k-1}^i)_{i=1}^n$, having distribution $\widetilde{\mathcal{S}}_k(\cdot)$ on $\{1,\ldots,n\}$, whose probabilities are proportional to
\begin{align}
& \widetilde{\mathcal{S}}_k(S_k = s) \propto \int_{[0,1]^m} \mathcal{U}(du) \nonumber \\
 & \quad \cdot \int_{\mathbb{X}} \psi_k(\mathscr{L}_{k-1}^{r_{k-1}^s(u)}, x_k) q_k(d x_k \, | \, \mathscr{L}_{k-1}^{r_{k-1}^s(u)}). \label{eq:twisted_S}
\end{align}
Also introduce a distribution $\widetilde{\mathcal{U}}_{k-1}(\,\cdot\,| \, s)$ on $[0,1]^m$ given by
\begin{align}
& \widetilde{\mathcal{U}}_{k-1}(du \, | \, s) \propto \nonumber \\
& \quad  \mathcal{U}(du)\int_{\mathbb{X}} \psi_k(\mathscr{L}_{k-1}^{r^{s}_{k-1}(u)},x_k) q_k(dx_k\,|\,\mathscr{L}_{k-1}^{r_{k-1}^s(u)}).\label{eq:twisted_U}
\end{align}
Note that the distributions $\widetilde{\mathcal{S}}_k$ and  $\widetilde{\mathcal{U}}_{k-1}$ depend on the resampling method defined through the mapping $r$.
Details of how to sample from these distributions in the cases when $r$ corresponds to multinomial or systematic resampling are given in Sections \ref{sec:twisted_multinomial} and \ref{sec:twisted_systematic}.

\renewcommand{\figurename}{\textbf{Algorithm}}
\begin{figure}[!t]
\begin{algorithmic}[1]
\State Sample $S_0$ uniformly from $\{1,\dots,n\}$
\State Sample $\xi_0^{S_0}\sim \widetilde{q_0}(\cdot)$
\For{$i \neq S_0$}
\State Sample $\xi_0^{i}\sim q_0(\cdot)$
\EndFor
\For{$1 \leq i \leq n$}
\State Set $W^i_0 = g_0(y_0 \, | \, \xi_0^i)\mu_0(\xi_0^i)/q_0(\xi_0^i)$
\EndFor
\State Set $\widetilde{Z}_{0} = \dfrac{\sum_{i=1}^n W_0^i \int_{\mathbb{X}} \psi_0(x_0) q_0(d x_0)}{\sum_{i=1}^n \psi_0(\xi_0^i)} $
\For{$1 \leq k \leq t$}
\State Sample $S_k \sim \widetilde{\mathcal{S}}_k(\cdot )$
\State Sample $U_{k-1} \sim \widetilde{\mathcal{U}}_{k-1}(\,\cdot\,| \,S_k)$
\State Set $A_{k-1}=r(U_{k-1},W_{k-1})$
\State Sample $\xi_k^{S_k} \sim \widetilde{q}_k(\cdot\, | \, \mathscr{L}_{k-1}^{A_{k-1}^{S_k}})$
\For{$ i \neq S_k$}
\State Sample $\xi_k^i \sim q_k(\cdot\, | \, \mathscr{L}_{k-1}^{A_{k-1}^i})$
\EndFor
\For{$ 1 \leq i \leq n$}
\State Set $W_k^i = \dfrac{g_k(y_{k}\, | \, \xi_{k}^i) f_k(\xi_{k}^i \, | \, \xi_{k-1}^{A_{k-1}^i})}{q_k(\xi_k^i \, | \, \mathscr{L}_{k-1}^{A_{k-1}^i})}$
\State Set $\widetilde{V}_{k-1}^i = \int_{\mathbb{X}} \psi_{k} (\mathscr{L}_{k-1}^{i},x)q_{k}(dx\,|\,\mathscr{L}_{k-1}^{i})$
\State Set $\widetilde{W}_{k-1}^i = W_{k-1}^i \widetilde{V}_{k-1}^i$
\EndFor
\State Set $\widetilde{Z}_{k} = \widetilde{Z}_{k-1} \dfrac{\sum_{i=1}^n W_k^i}{\sum_{i=1}^n W_{k-1}^i} \dfrac{\sum_{i=1}^n \widetilde{W}_{k-1}^i}{\sum_{i=1}^n \psi_k (\mathscr{L}_k^i)} $
\EndFor
\end{algorithmic}
\caption{Twisted particle filter}
\label{alg:twisted_pf}
\end{figure}

Our first main result, Theorem \ref{thm:twisted_unbiased}, establishes that Algorithm \ref{alg:twisted_pf} indeed samples from (\ref{eq:particle_filter_t0})--(\ref{eq:particle_filter_tk}) and delivers unbiased estimates of $p(y_{0:k})$, which justifies its use within Algorithm \ref{alg:pmmh}. The proof is given in Appendix \ref{app:proof_theorem1}.

\begin{theorem}\label{thm:twisted_unbiased}
The random variables $\xi_0$ and $(\xi_k, U_{k-1})_{k\geq 1}$ sampled using Algorithm \ref{alg:twisted_pf} are drawn from (\ref{eq:particle_filter_t0})--(\ref{eq:particle_filter_tk}). Furthermore, if Assumption \ref{ass:resampling} holds, then for each $k \geq 0$,
\begin{equation}
\widetilde{\mathbb{E}}[\widetilde{Z}_k]=\mathbb{E}[Z_k]=p(y_{0:k}), \label{eq:twisted_unbiased}
\end{equation}
where $\widetilde{\mathbb{E}}$ (resp. $\mathbb{E}$) denotes expectation w.r.t. \eqref{eq:particle_filter_t0}--\eqref{eq:particle_filter_tk} (resp. \eqref{eq:particle_filter_0}--\eqref{eq:particle_filter_k}).
\end{theorem}

Theorem \ref{thm:optimal} identifies the choice of the functions $(\psi_k)_{0\leq k \leq t}$ which are
ideal for estimating $p(y_{0:t})$. The proof is given in Appendix \ref{app:proof_theorem2}.
\begin{theorem}\label{thm:optimal}
If we choose
\begin{align}
\psi_{k}(x_{0:k}) & = \left\{ \begin{array}{ll}  \dfrac{\mu_0(x_0)p(y_{0:t} \, | \, x_0)}{q_0(x_0)}, & k= 0, \\ \dfrac{f_{k}(x_{k} \, | \, x_{k-1})p(y_{k:t} \, | \, x_{k})}{q_k(x_{k} \, | \,x_{0:k-1})}, & 1 \leq k \leq t, \end{array} \right. \label{eq:twisted_optimal_phi}
\end{align}
then $\widetilde{Z}_t = p(y_{0:t})$.
\end{theorem}

The choice of $\psi_k$ identified in  (\ref{eq:twisted_optimal_phi}) is of course not usually available in practice, but Theorem \ref{thm:optimal} motivates us to consider twisting functions of the form
\begin{align} \label{eq:approx_twisting}
\psi_{k,l}(x_{0:k}) & = \left\{ \begin{array}{cc} \dfrac{\mu_0(x_0)\phi_{0,l}(x_0)}{q_0(x_0)}, & k = 0, \\ \dfrac{f_{k}(x_k \, | \, x_{k-1}) \phi_{k,l}(x_{0:k})}{q_k(x_k \, | \, x_{0:k-1})}, & 1 \leq k \leq t,
\end{array} \right.
\end{align}
where the functions $\phi_{k,l}\colon \mathbb{X}^{k+1} \to [0,1]$ are chosen to approximate $p(y_{k:k+l} \, | \, x_k)$, possibly also depending on $\mathscr{F}_{k-1}$, and $l$ is a positive integer parameter such that $0 \leq l \leq t-k$, which specifies how many future measurements are used in the twisting function.
Devising such approximations is the subject of Section \ref{sec:twisted_for_Gaussian}.

We conclude Section \ref{sec:twisted_pf} by showing how to sample $S_k$ and $U_{k-1}$ on lines 11 and 12 in Algorithm \ref{alg:twisted_pf}.

\subsection{Twisted multinomial resampling} \label{sec:twisted_multinomial}
In this case $m=n$, and using definition (\ref{eq:multinomial}) for $r_{k-1}$, it is easily checked that the probabilities $\widetilde{\mathcal{S}}_k(S_k = s)$ in  (\ref{eq:twisted_S}) are independent of the value $s$, i.e. $\widetilde{\mathcal{S}}_k$ is the uniform distribution over $\{1,\ldots,n\}$.

The density function corresponding to (\ref{eq:twisted_U}) can be written as
\begin{align*}
& \widetilde{\mathcal{U}}_{k-1}(u \, | \, s) \\
& \propto \mathbb{I}_{[0,1]}(u^s)\int_{\mathbb{X}} \psi_k(\mathscr{L}_{k-1}^{r^{s}_{k-1}(u^s)},x_k) q_k(dx_k\,|\,\mathscr{L}_{k-1}^{r_{k-1}^s(u^s)}) \\
&\quad  \cdot \prod_{i \neq s} \mathbb{I}_{[0,1]}(u^i) \\
& = \left[\sum_{j=1}^n \mathbb{I}_{(d^{j-1}, d^j]}(u^s)\tilde{v}_{k-1}^j\right] \prod_{i \neq s} \mathbb{I}_{[0,1]}(u^i),
\end{align*}
where the equality uses  (\ref{eq:multinomial}), and  $d_{k-1}^0 = \ 0$, $d_{k-1}^j = \sum_{i=1}^j w_{k-1}^i / \sum_{i=1}^n w_{k-1}^i$ for $1\leq j \leq n$, for any set $\mathcal{I}$, $\mathbb{I}_{\mathcal{I}}(u)=1$, when $u \in \mathcal{I}$ and zero otherwise, and the terms $\tilde{v}_{k-1}^j$ are given by  (\ref{eq:twisted_v}).

We therefore have the following procedure for sampling $S_k$ and $U_{k-1}$ from $\widetilde{\mathcal{S}}_{k}(\cdot)$ and $\widetilde{\mathcal{U}}_{k-1}(\cdot \, | \, s)$ respectively:
\begin{enumerate}
\item Sample $S_k$ uniformly from $\{1,\ldots,n\}$
\item Sample index $J_{k-1}$ from the discrete distribution on $\{1,\ldots,n\}$ such that the probability that $J_{k-1} = j$ is proportional to
\begin{align*}
 & \int_{[0,1]}\mathbb{I}_{(d_{k-1}^{j-1}, d_{k-1}^j]}(u^s)\, du^s \, \tilde{v}_{k-1}^j  = w_{k-1}^j \tilde{v}_{k-1}^j
\end{align*}
\item Sample $U_{k-1}^{S_k}$ from the uniform distribution on $(d_{k-1}^{J_{k-1}-1}, d_{k-1}^{J_{k-1}}]$ and for each $i \neq S_k$, $U_{k-1}^i$ from the uniform distribution on $[0,1]$
\end{enumerate}

\subsection{Twisted systematic resampling} \label{sec:twisted_systematic}
In this case we have $m=1$, and using definition (\ref{eq:systematic}) for $r_{k-1}$, the probabilities in  (\ref{eq:twisted_S}) are
\begin{align}
& \widetilde{\mathcal{S}}_k(S_k = s) \nonumber \\
 & \propto \int_{[0,1]} \mathcal{U}(du) \int_{\mathbb{X}} q_k(d x_k \, | \, \mathscr{L}_{k-1}^{r_{k-1}^s(u)}) \psi_k(\mathscr{L}_{k-1}^{r_{k-1}^s(u)}, x_k) \nonumber \\
 & = \sum_{j=1}^n \int_{[0,1]} \mathbb{I}_{\mathcal{I}^{s,j}_{k-1}}(u) \, du \int_{\mathbb{X}} q_k(d x_k \, | \, \mathscr{L}_{k-1}^j) \psi_k(\mathscr{L}_{k-1}^j, x_k) \nonumber \\
 & = \sum_{\{j \, | \, \mathcal{I}_{k-1}^{j,s} \neq \emptyset\}} \left[\min(nd_{k-1}^j-s+1,1) \right.  \nonumber \\
& \left. \quad \quad \quad \quad  \quad \quad - \max(nd_{k-1}^{j-1}-s+1,0)\right] \tilde{v}_{k-1}^j, \label{eq:systematic_S}
\end{align}
where the first equality follows from (\ref{eq:systematic}), and $\mathcal{I}^{s,j}_{k-1} = (nd_{k-1}^{j-1}-s+1, nd_{k-1}^j-s+1] \cap [0,1]$ and $(d_{k-1}^j)_{j=0}^n$ are defined as in the twisted multinomial resampling.

The probability density function corresponding to (\ref{eq:twisted_U}) can be written as
\begin{align*}
& \widetilde{\mathcal{U}}_{k-1}(u \, | \, s) \\
& \propto \mathbb{I}_{[0,1]}(u)\int_{\mathbb{X}} \psi_k(\mathscr{L}_{k-1}^{r^{s}_{k-1}(u)},x_k) q_k(dx_k\,|\,\mathscr{L}_{k-1}^{r_{k-1}^s(u)}) \\
& = \sum_{j=1}^n \mathbb{I}_{\mathcal{I}_{k-1}^{s,j}}(u) \tilde{v}_{k-1}^j,
\end{align*}
where the equality follows from (\ref{eq:systematic}).

This leads to the following procedure for sampling $S_k$ and $U_{k-1}$ from $\widetilde{\mathcal{S}}_k(\cdot)$ and $\widetilde{\mathcal{U}}_{k-1}(\cdot \, | \, s)$ respectively:
\begin{enumerate}
\item Sample $S_k$ from a distribution over $\{1,\ldots,n\}$ with probabilities given by  (\ref{eq:systematic_S})
\item Sample index $J_{k-1}$ from the discrete distribution on $\{1,\ldots,n\}$ such that the probability that  $J_{k-1}=j$ is proportional to
\begin{align*}
&  \int_{[0,1]}\mathbb{I}_{\mathcal{I}^{S_k,j}_{k-1}}(u)\, du \, \tilde{v}_{k-1}^j \\
& = \left[\min(nd_{k-1}^j-S_k+1,1) \right.  \\
& \left. \quad \quad - \max(nd_{k-1}^{j-1}-S_k+1,0)\right] \tilde{v}_{k-1}^j ,
\end{align*}
if $\mathcal{I}^{S_k,j}_{k-1} \neq \emptyset$, and otherwise  the probability that $J_{k-1}=j$ is zero.
\item Sample $U_{k-1}$ from the uniform distribution on $\mathcal{I}^{S_k,J_{k-1}}_{k-1}$
\end{enumerate}

\subsection{Complexity of twisted resampling methods}%new in the revision

Twisted multinomial resampling involves sampling 2 times from a discrete distribution with $n$ elements and $n$ times from a continuous uniform distribution, and can be implemented in $\mathcal{O}(n)$ time.

Twisted systematic resampling involves sampling two times from a discrete distribution with $n$ elements and one time from a continuous uniform distribution, and can be implemented in $\mathcal{O}(n)$ time.
Compared to twisted multinomial resampling, some computation time is saved since only one draw from the continuous uniform distribution is needed.
However, the computation of the probabilities for the discrete distributions is computationally more involved for the twisted systematic resampling.

The overall complexity of Algorithm \ref{alg:twisted_pf} depends on the specific nature of the twisting function and how it is computed. This is a problem-specific issue, which we discuss in the context of a particular family of models and twisting functions in Section \ref{sec:complexity_gaussian}.

\section{Twisted particle filters for Gaussian state-space models}
\label{sec:twisted_for_Gaussian}

In this section, we present methods for approximating the optimal twisting function in Gaussian state-space models with $\mathbb{X}=\mathbb{R}^{d_x}$, $\mathbb{Y}=\mathbb{R}^{d_y}$ and
\begin{subequations}
\label{eq:Gaussian_model}
\begin{align}
\mu_0(\cdot) &= \mathcal{N}(\cdot \, | \, \nu_0, \mathbf{P}_0), \\
f_{k}(\cdot \, | \, x_{k-1}) & = \mathcal{N}(\cdot \, | \, c_{k-1}(x_{k-1}), \mathbf{Q}_{k-1}), \quad k \geq 1, \\
g_k(\cdot \, | \, x_k) & = \mathcal{N}(\cdot \, | \, h_{k}(x_k), \mathbf{R}_{k}), \quad k \geq 0,
\end{align}
\end{subequations}
where $\mathcal{N}(\cdot \, | \, \nu, \mathbf{P})$ denotes a Gaussian distribution with mean vector $\nu$ and covariance matrix $\mathbf{P}$.
The mean functions $c_{k-1}(x_{k-1})$ and $h_{k}(x_k)$ can be nonlinear functions of the state vector.

To use the twisted particle filter in practice, we need to evaluate the integrals in  (\ref{eq:twisted_v}) and sample from the twisted distributions given by  (\ref{eq:twisted_q}).
For the Gaussian model, we choose an exponential form for the function $\phi_{k,l}$ in  (\ref{eq:approx_twisting}), given by
\begin{align}
\phi_{k,l}(x_{0:k})  =  \alpha_{k,l} \exp\left\{ - \frac{1}{2} x_k^T \boldsymbol{\Gamma}_{k,l} x_k + x_k^T\beta_{k,l} \right\},
\label{eq:twist_phi}
\end{align}
where $\alpha_{k,l} \equiv \alpha_{k,l}(x_{0:k-1}) \in \mathbb{R}^+$, $\beta_{k,l} \equiv \beta_{k,l}(x_{0:k-1}) \in \mathbb{R}^{d_x}$ and $\boldsymbol{\Gamma}_{k,l} \equiv \boldsymbol{\Gamma}_{k,l}(x_{0:k-1}) \in \mathbb{R}^{d_x \times d_x}$ are parameters, possibly depending on $\mathscr{F}_{k-1}$ and any number of measurements.
For $k \geq 1$, we use shorthand notation $\alpha_{k,l}^i = \alpha_{k,l}(\mathscr{L}_{k-1}^i)$, $\beta_{k,l}^i = \beta_{k,l}(\mathscr{L}_{k-1}^i)$ and $\boldsymbol{\Gamma}_{k,l}^i = \boldsymbol{\Gamma}_{k,l}(\mathscr{L}_{k-1}^i)$.
Methods for computing these parameters are considered in Sections \ref{sec:local_linearization} and \ref{sec:mode_linearization}.

With twisting function given by  (\ref{eq:approx_twisting}) and (\ref{eq:twist_phi}), we have $\widetilde{q}_0(\cdot) = \mathcal{N}(\cdot \, | \, \mu_{0,l}, \boldsymbol{\Sigma}_{0,l})$, where
\begin{subequations}
\begin{align}
\mu_{0,l}  & =  \boldsymbol{\Sigma}_{0,l} \left(\mathbf{P}_0^{-1} \nu_0 + \beta_{0,l} \right), \label{eq:mu0_Gaussian} \\
 \boldsymbol{\Sigma}_{0,l}  & =  \left(\mathbf{P}_0^{-1} + \boldsymbol{\Gamma}_{0,l} \right)^{-1}. \label{eq:Sigma0_Gaussian}
 \end{align}
 \end{subequations}
For $k \geq 1$ and $1\leq i \leq n$, we have $\widetilde{q}_k(\cdot \, | \, \mathscr{L}_{k-1}^i) = \mathcal{N}(\cdot \, | \, \mu_{k,l}^i, \mathbf{\Sigma}_{k,l}^i)$, where
\begin{subequations}
\begin{align}
\mu_{k,l}^i &= \boldsymbol{\Sigma}_{k,l}^i \left(\mathbf{Q}_{k-1}^{-1}c_{k-1}(\xi_{k-1}^i) + \beta_{k,l}^i\right),  \label{eq:mu_Gaussian} \\
\boldsymbol{\Sigma}_{k,l}^i & =  \left(\mathbf{Q}_{k-1}^{-1} + \boldsymbol{\Gamma}^i_{k,l} \right)^{-1}. \label{eq:Sigma_Gaussian}
\end{align}
\end{subequations}

The initial likelihood estimate in the twisted particle filter is now given by
\begin{equation}
\widetilde{Z}_{0} = \left[\frac{\alpha_{0,l}|\boldsymbol{\Sigma}_{0,l}|^{1/2}}{|\mathbf{P}_0|^{1/2}}\frac{\exp\left\{ \frac{1}{2}\mu_{0,l} ^T\boldsymbol{\Sigma}_{0,l}^{-1} \mu_{0,l} \right\}}{\exp\left\{ \frac{1}{2} \nu_0^T \mathbf{P}_0^{-1} \nu_0 \right\}}\right]\frac{\sum_{i=1}^n W_0^i}{\sum_{i=1}^n \psi_0(\xi_0^i)},
\label{eq:twisted_Z0_gaussian}
\end{equation}
where $| \mathbf{P} |$ denotes the determinant of a matrix $\mathbf{P}$.
The integral in  (\ref{eq:twisted_v}) can be computed to give
\begin{align}
 \widetilde{V}_{k}^i & = \alpha^i_{k+1,l}  \frac{|\boldsymbol{\Sigma}^i_{k+1,l}|^{1/2}}{|\mathbf{Q}_k|^{1/2}} \nonumber \\
  & \quad \quad \cdot \frac{\exp\left\{ \frac{1}{2}(\mu^{i}_{k+1,l})^T \left(\boldsymbol{\Sigma}^i_{k+1,l}\right)^{-1} \mu^{i}_{k+1,l} \right\}}{\exp\left\{ \frac{1}{2} c_k(\xi_{k}^i)^T \mathbf{Q}_k^{-1} c_k(\xi_{k}^i)\ \right\} },
\label{eq:twisted_V_gaussian}
\end{align}
for $k \geq 0$ and $1 \leq i \leq n$.

The resulting algorithm for the twisted particle filter for Gaussian state-space models is given in Algorithm \ref{alg:twisted_pf_gaussian}.
We conclude Section \ref{sec:twisted_for_Gaussian} by presenting two methods for computing the twisting function parameters on lines 1 and 13. Whilst we focus on the case of Gaussian disturbances, one could follow an almost identical approach to constructing a twisting function for a model in which the disturbances are non-Gaussian, but of known mean and covariance. In particular, one replaces respectively $\mathbf{Q}_{k-1}$ and $\mathbf{R}_k$ by the conditional covariances of $X_k|x_{k-1}$ and $Y_k|x_k$, and $c_{k-1}(x_{k-1})$ and $h_k(x_k)$ by the conditional means of $X_k|x_{k-1}$ and $Y_k|x_k$. Exponential-family disturbances could be treated with the kind of techniques explored in \cite{hlinka2012likelihood}.

\begin{figure}[!t]
\begin{algorithmic}[1]
\State Set $(\alpha_{0,l}, \beta_{0,l},\boldsymbol{\Gamma}_{0,l})$ using Algorithm \ref{alg:local_linearization_t0} or \ref{alg:mode_linearization}
\State Set $\mu_{0,l}$ and $\boldsymbol{\Sigma}_{0,l}$ using  (\ref{eq:mu0_Gaussian})--(\ref{eq:Sigma0_Gaussian})
\State Sample $S_0$ uniformly from $\{1,\dots,n\}$
\State Sample $\xi_0^{S_0}\sim \mathcal{N}(\cdot \, | \, \mu_{0,l}, \boldsymbol{\Sigma}_{0,l})$
\For{$i \neq S_0$}
\State Sample $\xi_0^{i}\sim q_0(\cdot)$
\EndFor
\For{$1 \leq i \leq n$}
\State Set $W^i_0 = g(y_0 \, | \, \xi_0^i)\mu_0(\xi_0^i)/q_0(\xi_0^i)$
\EndFor
\State Set $\widetilde{Z}_{0}$ using  (\ref{eq:twisted_Z0_gaussian})
\For{$1 \leq k \leq t$}
\State Set $(\alpha_{k,l}^i, \beta_{k,l}^i,\boldsymbol{\Gamma}_{k,l}^i)_{i=1}^n$ using Algorithm \ref{alg:local_linearization_tk} or \ref{alg:mode_linearization}
\For{$1 \leq i \leq n$}
\State Set $\mu_{k,l}^i$ and $\boldsymbol{\Sigma}_{k,l}^i$ using  (\ref{eq:mu_Gaussian})--(\ref{eq:Sigma_Gaussian})
\EndFor
\State Sample $S_k \sim \widetilde{\mathcal{S}}_k(\cdot)$
\State Sample $U_{k-1} \sim \widetilde{\mathcal{U}}_{k-1}(\,\cdot\,| \,S_k)$
\State Set $A_{k-1}=r(U_{k-1},W_{k-1})$
\State Sample $\xi_k^{S_k} \sim \mathcal{N}(\cdot\, | \, \mu_{k,l}^{A_{k-1}^{S_k}}, \boldsymbol{\Sigma}_{k,l}^{A_{k-1}^{S_k}})$
\For{$ i \neq S_k$}
\State Sample $\xi_k^i \sim q_k(\cdot\, | \, \mathscr{L}_{k-1}^{A_{k-1}^i})$
\EndFor
\For{$ 1 \leq i \leq n$}
\State Set $W_k^i = \dfrac{g(y_{k}\, | \, \xi_{k}^i) f(\xi_{k}^i \, | \, \xi_{k-1}^{A_{k-1}^i})}{q_k(\xi_k^i \, | \, \mathscr{L}_{k-1}^{A_{k-1}^i})}$
\State Set $\widetilde{V}_{k-1}^i$ using  (\ref{eq:twisted_V_gaussian}) and $\widetilde{W}_{k-1}^i = W_{k-1}^i \widetilde{V}_{k-1}^i$
\EndFor
\State Set $\widetilde{Z}_{k} = \widetilde{Z}_{k-1} \dfrac{\sum_{i=1}^n W_k^i}{\sum_{i=1}^n W_{k-1}^i} \dfrac{\sum_{i=1}^n \widetilde{W}_{k-1}^i}{\sum_{i=1}^n \psi_k (\mathscr{L}_k^i)} $
\EndFor
\end{algorithmic}
\caption{Twisted particle filter for Gaussian model}
\label{alg:twisted_pf_gaussian}
\end{figure}

\subsection{Twisting function using local linearization}
\label{sec:local_linearization}

For a linear Gaussian model the term $p(y_{k:k+l} \, | \, x_k)$, as a function of $x_k$, is exactly of the exponential form in  (\ref{eq:twist_phi}).
For a nonlinear Gaussian model, we can therefore compute an approximation of $p(y_{k:k+l} \, | \, x_k)$ by considering linearized transition and measurement functions.

We propose to use a local Taylor series based linearization using the extended Kalman filter (EKF).
The local linearization method for computing the twisting function parameters $\alpha_{k,l}$, $\beta_{k,l}$ and $\boldsymbol{\Gamma}_{k,l}$ is summarized in Algorithms \ref{alg:local_linearization_t0} and \ref{alg:local_linearization_tk} and details are given in the following equations.

\begin{figure}[!t]
\begin{algorithmic}[1]
\State Set $\hat{x}_{0}^- = \nu_0$ and $\hat{\mathbf{P}}_0^- = \mathbf{P}_0$
\State Set $\hat{x}_0$, $\hat{\mathbf{P}}_0$, $\mathbf{H}_0$ and $\hat{h}_0$ using  (\ref{eq:ekf_update})-(\ref{eq:Hk_re})
\State Set $(\alpha_{0,0}$, $\beta_{0,0}$, $\boldsymbol{\Gamma}_{0,0})$ using  (\ref{eq:par_init})
\For{$1\leq s \leq l$}
\State Set $\hat{x}_{s}$, $\hat{\mathbf{P}}_{s}$, $\mathbf{C}_{s-1}$, $\hat{c}_{s-1}$, $\mathbf{H}_{s}$ and $\hat{h}_{s}$ using  (\ref{eq:ekf_prediction})-(\ref{eq:Hk_re})
\State Set $(\alpha_{0,s}$, $\beta_{0,s}$, $\boldsymbol{\Gamma}_{0,s})$ using  (\ref{eq:par_rec})--(\ref{eq:auxiliary})
\EndFor
\State Return $(\alpha_{0,l}, \beta_{0,l}, \boldsymbol{\Gamma}_{0,l})$
\end{algorithmic}
\caption{Twisting function parameters for $k=0$ using EKF linearization}
\label{alg:local_linearization_t0}
\end{figure}

\begin{figure}[!t]
\begin{algorithmic}[1]
\For{$1\leq i \leq n$}
\State Set $\hat{x}_{k-1}^i = \xi_{k-1}^i$ and $\hat{\mathbf{P}}_{k-1}^i = \boldsymbol{0}$
\State Set $\hat{x}_k^i$, $\hat{\mathbf{P}}_k^i$, $\mathbf{H}_k^i$ and $\hat{h}_k^i$ using  (\ref{eq:ekf_prediction})--(\ref{eq:Hk_re})
\State Set $(\alpha_{k,0}^i$, $\beta_{k,0}^i$, $\boldsymbol{\Gamma}_{k,0}^i)$ using  (\ref{eq:par_init})
\For{$1\leq s \leq l$}
\State Set $\hat{x}_{k+s}^i$, $\hat{\mathbf{P}}_{k+s}^i$, $\mathbf{C}_{k+s-1}^i$, $\hat{c}_{k+s-1}^i$, $\mathbf{H}_{k+s}^i$, $\hat{h}_{k+s}^i$ using  (\ref{eq:ekf_prediction})--(\ref{eq:Hk_re})
\State Set $(\alpha_{k,s}^i$, $\beta_{k,s}^i$, $\boldsymbol{\Gamma}_{k,s}^i)$ using  (\ref{eq:par_rec})--(\ref{eq:auxiliary})
\EndFor
\EndFor
\State Return $(\alpha_{k,l}^i, \beta_{k,l}^i, \boldsymbol{\Gamma}_{k,l}^i)_{i=1}^n$
\end{algorithmic}
\caption{Twisting function parameters for $k \geq 1$ using local EKF linearization}
\label{alg:local_linearization_tk}
\end{figure}

We first present the equations for computing the linearized transition functions $c_{k+s-1}(x_{k+s-1})  \approx \mathbf{C}_{k+s-1}x_{k+s-1} + \hat{c}_{k+s-1}$ and linearized measurement functions $h_{k+s}(x_{k+s})  \approx \mathbf{H}_{k+s} x_{k+s} + \hat{h}_{k+s}$ for $0 \leq s \leq l$ using the EKF local linearization.

For $k \geq 1$ and $1 \leq i \leq n$, the EKF algorithm is initialized with $\hat{x}_{k-1} = \xi^i_{k-1}$ and $\hat{\mathbf{P}}_{k-1} = \boldsymbol{0}$.
For $0 \leq s \leq l$, we recursively compute $\mathbf{C}_{k+s-1}$, $\hat{c}_{k+s-1}$, $\mathbf{H}_{k+s}$ and $\hat{h}_{k+s}$ by first linearizing the transition function using the EKF prediction step equations
\begin{subequations}
\label{eq:ekf_prediction}
\begin{align}
\hat{x}_{k+s}^- & = c_{k+s-1}(\hat{x}_{k+s-1}),  \\
\mathbf{C}_{k+s-1} & = \left[ \frac{\partial }{\partial x} c_{k+s-1}(x) \right]_{x = \hat{x}_{k+s-1}}, \\
\hat{c}_{k+s-1}  & =c_{k+s-1}(\hat{x}_{k+s-1}) - \mathbf{C}_{k+s-1}\hat{x}_{k+s-1}, \label{eq:hatc} \\
\hat{\mathbf{P}}^-_{k+s} & = \mathbf{C}_{k+s-1} \hat{\mathbf{P}}_{k+s-1} \mathbf{C}_{k+s-1}^T + \mathbf{Q}_{k+s-1},
\end{align}
\end{subequations}
where for a vector-valued function $c$, $\left[ \frac{\partial }{\partial x} c(x) \right]_{x = \hat{x}}$ denotes the Jacobian matrix, evaluated at the point $x=\hat{x}$.

The linearization for the measurement function is obtained by first computing the EKF update step equations
\begin{subequations}
\label{eq:ekf_update}
\begin{align}
\mathbf{H}_{k+s}^- & = \left[ \frac{\partial }{\partial x} h_{k+s}(x) \right]_{x = \hat{x}_{k+s}^-},  \label{eq:Hk}\\
\mathbf{S}_{k+s}^- & = \mathbf{H}_{k+s}^- \hat{\mathbf{P}}_{k+s}^- \left(\mathbf{H}_{k+s}^-\right)^T + \mathbf{R}_{k+s}, \label{eq:Sk}\\
\mathbf{G}_{k+s}^- &  = \hat{\mathbf{P}}_{k+s}^- \left(\mathbf{H}^-_{k+s}\right)^T \left(\mathbf{S}^-_{k+s}\right)^{-1}, \label{eq:Gk} \\
\hat{x}_{k+s} & = \hat{x}_{k+s}^- + \mathbf{G}^-_{k+s}(y_{k+s} - h_{k+s}(\hat{x}_{k+s}^-)), \\
\hat{\mathbf{P}}_{k+s} & = \hat{\mathbf{P}}_{k+s}^- - \mathbf{G}^-_{k+s} \mathbf{S}_{k+s}^- \left(\mathbf{G}_{k+s}^-\right)^T,
\end{align}
\end{subequations}
and then relinearizing w.r.t. $\hat{x}_{k+s}$:
\begin{subequations}
\label{eq:Hk_re}
\begin{align}
\mathbf{H}_{k+s} & = \left[ \frac{\partial }{\partial x} h_{k+s}(x) \right]_{x = \hat{x}_{k+s}},  \\
\hat{h}_{k+s} &  =  h_{k+s}(\hat{x}_{k+s}) - \mathbf{H}_{k+s}  \hat{x}_{k+s}.
\end{align}
\end{subequations}
For $k=0$, the EKF algorithm is initialized using $\hat{x}_0^- = \nu_0$ and $\hat{\mathbf{P}}_0^- = \mathbf{P}_0$ and the recursion is started from the update step (\ref{eq:ekf_update}).

The parameters $\alpha_{k,l}$, $\beta_{k,l}$ and $\boldsymbol{\Gamma}_{k,l}$ in  (\ref{eq:twist_phi}) can be then computed recursively using the following equations.
The parameters are initialized with
\begin{subequations}
\label{eq:par_init}
\begin{align}
\alpha_{k,0} & = \frac{\exp\{ -\frac{1}{2} (y_k-\hat{h}_k)^T \mathbf{R}_k^{-1}(y_k-\hat{h}_k)\}}{|2 \pi \mathbf{R}_k|^{1/2}}, \\
 \beta_{k,0} & = \mathbf{H}_k^T \mathbf{R}_k^{-1}(y_k-\hat{h}_k), \\
\boldsymbol{\Gamma}_{k,0} & = \mathbf{H}_k^T \mathbf{R}_k^{-1} \mathbf{H}_k.
\end{align}
\end{subequations}
Recursive updates for $1 \leq s \leq l$ are given by
\begin{subequations}
\label{eq:par_rec}
\begin{align}
\alpha_{k,s} & = \alpha_{k,s-1} \frac{\exp\left\{ -\frac{1}{2}\epsilon_{k+s}^T \mathbf{S}_{k+s}^{-1} \epsilon_{k+s} \right\} }{| \mathbf{S}_{k+s} |^{1/2}}, \\
\beta_{k,s} & = \beta_{k,s-1} + \mathbf{D}_{k+s}^T \mathbf{H}_{k+s}^T \mathbf{S}_{k+s}^{-1}\epsilon_{k+s}, \\
\boldsymbol{\Gamma}_{k,s} & = \boldsymbol{\Gamma}_{k,s-1} + \mathbf{D}_{k+s}^T \mathbf{H}_{k+s}^T \mathbf{S}_{k+s}^{-1} \mathbf{H}_{k+s} \mathbf{D}_{k+s},
\end{align}
\end{subequations}
where
\begin{subequations}
\begin{align}
\epsilon_{k+s} & = y_{k+s}-\hat{h}_{k+s} - \mathbf{H}_{k+s} v_{k+s}, \\
\mathbf{S}_{k+s} & = \mathbf{H}_{k+s} \mathbf{K}_{k+s}\mathbf{H}_{k+s}^T + \mathbf{R}_{k+s}, \label{eq:Sk_re}\\
\mathbf{G}_{k+s} &  = \mathbf{K}_{k+s} \mathbf{H}_{k+s}^T \mathbf{S}_{k+s}^{-1}, \label{eq:Gk_re}
\end{align}
\end{subequations}
and the variables $\mathbf{D}_{k+s}$, $\mathbf{K}_{k+s}$ and $v_{k+s}$ are initialized with $\mathbf{D}_{k+1} = \mathbf{C}_k$, $\mathbf{K}_{k+1} = \mathbf{Q}_{k}$ and $v_{k+1} = \hat{c}_k$, and then recursively computed for $2 \leq s \leq l$ using
\begin{subequations}
\label{eq:auxiliary}
\begin{align}
\mathbf{D}_{k+s} & = (\mathbf{C}_{k+s-1} - \mathbf{C}_{k+s-1} \mathbf{G}_{k+s-1}\mathbf{H}_{k+s-1})\mathbf{D}_{k+s-1}, \label{eq:D_update} \\
\mathbf{K}_{k+s} & = \mathbf{C}_{k+s-1} \left(\mathbf{K}_{k+s-1}  \right. \nonumber \\
 & \quad \quad \left. - \mathbf{G}_{k+s-1} \mathbf{S}_{k+s-1} \mathbf{G}_{k+s-1}^T\right)\mathbf{C}_{k+s-1}^T + \mathbf{Q}_{k+s-1}, \label{eq:K_update} \\
 v_{k+s} & = \mathbf{C}_{k+s-1} \left[v_{k+s-1} +\mathbf{G}_{k+s-1}\epsilon_{k+s-1}\right] + \hat{c}_{k+s-1}.  \label{eq:v_update}
\end{align}
\end{subequations}

The computational complexity of Algorithms \ref{alg:local_linearization_t0} and \ref{alg:local_linearization_tk} are $\mathcal{O}(nl))$.
To reduce computational time, it is possible to leave out the relinearization of the measurement function and set $\mathbf{H}_{k+s} = \mathbf{H}_{k+s}^-$ and $\hat{h}_{k+s}  =  h_{k+s}(\hat{x}_{k+s}^-) + \mathbf{H}_{k+s}^-  \hat{x}_{k+s}^-$.
We then have $\mathbf{S}_{k+s} = \mathbf{S}_{k+s}^-$ and $\mathbf{G}_{k+s} = \mathbf{G}_{k+s}^-$, and we therefore do not need to evaluate  (\ref{eq:Sk_re}), (\ref{eq:Gk_re}) and (\ref{eq:K_update}) when computing the parameters $\alpha_{k,l}$, $\beta_{k,l}$ and $\boldsymbol{\Gamma}_{k,l}$.
However, in our experiments, the increase in performance when using relinearization was found to clearly outweigh the increase in computational time.

\subsection{Twisting function using linearization around the mode}
\label{sec:mode_linearization}

The local linearization approximation requires running the EKF algorithm separately for each particle to obtain the corresponding twisting function parameters.
This is computationally heavy and can make the local linearization approach too slow in practice.
Computation time can be significantly reduced if we can make some assumptions about the form of $p(y_{k:k+l} \, | \, x_k)$.

The simplest case is when $p(y_{k:k+l} \, | \, x_k)$ can be assumed to be roughly symmetric and unimodal.
A global approximation can be then obtained by computing the twisting function parameters using EKF linearization around the mode.
This method has computational complexity of $\mathcal{O}(l)$ and is summarized in Algorithm \ref{alg:mode_linearization}.

In practice, an approximation to the location of the mode can be obtained by using a Gaussian smoother initialized from some distribution over $x_k$ set for example as some function of the particles $(\xi_{k-1}^i)_{i=1}^n$, to approximate the mean of $p(x_k \, | \, y_{k:k+l})$.
We can then take the smoothed mean as an approximation for the mode.
More accurate approximation of the mode can be obtained by targeting $\log p(y_{k:k+l} \, | \, x_k)$ directly and using an iterative optimization method.

For multimodal $p(y_{k:k+l} \, | \, x_k)$, the linearization could be done separately for all the modes and then combined into a mixture of exponential terms of the form in  (\ref{eq:twist_phi}) (see \cite{ali-loytty2007gaussian} where a similar approach is used to approximate multimodal likelihoods in Gaussian mixture filters).

\begin{figure}[!t]
\begin{algorithmic}[1]
\State Set $\hat{x}_{k} \approx \arg \max_{x_k} p(y_{k:k+l} \, | \, x_k)$ and $\mathbf{P}_k= \boldsymbol{0}$
\State Set $\mathbf{H}_k$ and $\hat{h}_k$ using  (\ref{eq:Hk_re})
\State Set $(\alpha_{k,0}$, $\beta_{k,0}$, $\boldsymbol{\Gamma}_{k,0})$ using  (\ref{eq:par_init})
\For{$1\leq s \leq l$}
\State Set $\hat{x}_{k+s}$, $\mathbf{P}_{k+s}$, $\mathbf{C}_{k+s-1}$, $\hat{c}_{k+s-1}$, $\mathbf{H}_{k+s}$ and $\hat{h}_{k+s}$ using  (\ref{eq:ekf_prediction})--(\ref{eq:Hk_re})
\State Set $(\alpha_{k,s}$, $\beta_{k,s}$, $\boldsymbol{\Gamma}_{k,s})$ using  (\ref{eq:par_rec})--(\ref{eq:auxiliary})
\EndFor
\State Set $\alpha_{k,l}^i = \alpha_{k,l}$, $\beta_{k,l}^i=\beta_{k,l}$ and $\boldsymbol{\Gamma}_{k,l}^i = \boldsymbol{\Gamma}_{k,l}$ for all $1 \leq i \leq n$
\State Return $(\alpha_{k,l}^i, \beta_{k,l}^i, \boldsymbol{\Gamma}_{k,l}^i)_{i=1}^n$
\end{algorithmic}
\caption{Twisting function parameters using EKF linearization around the mode of $p(y_{k:k+l} \, | \, x_k)$}
\label{alg:mode_linearization}
\end{figure}

\subsection{Complexity of twisted particle filters using linearization}\label{sec:complexity_gaussian}
First consider Algorithm \ref{alg:twisted_pf_gaussian} in the case that Algorithms \ref{alg:local_linearization_t0} and \ref{alg:local_linearization_tk}  are used at lines $1$ and $13$ respectively.
Algorithms \ref{alg:local_linearization_t0} and \ref{alg:local_linearization_tk} have computational complexity $\mathcal{O}(nl)$ and the full Algorithm \ref{alg:twisted_pf_gaussian} then scales as $\mathcal{O}(tnl)$.

Consider next using Algorithm \ref{alg:mode_linearization} at lines $1$ and $13$ in Algorithm \ref{alg:twisted_pf_gaussian}.
Algorithm \ref{alg:mode_linearization} has computational complexity $\mathcal{O}(l)$ and the overall complexity of Algorithm \ref{alg:twisted_pf_gaussian} then scales as $\mathcal{O}(t(n+l))$.

\section{Applications and numerical results}\label{sec:applications}

We provide here numerical examples to demonstrate the use of twisted particle filter and compare its performance against a particle filter in likelihood estimation and parameter inference using particle MCMC.

We consider the following particle filters:
\begin{itemize}
\item \textbf{BSPF:} bootstrap particle filter, i.e. $q_k=f_k$
\item \textbf{EKFPF:} particle filter in which $q_k$ is obtained by a standard EKF local-linearization of the importance distribution minimizing the conditional expectation of the importance weights -- see \cite{doucet2000sequential} for details.
\item \textbf{twisted-BSPF-local:} twisted version of BSPF using the EKF local linearization for the twisting function.
\item \textbf{twisted-EKFPF-local:} twisted version of EKFPF using the EKF local linearization for the twisting function.
\item \textbf{twisted-BSPF-mode:} computationally lighter alternative for the twisted-BSPF-local, where we use EKF linearization around an approximation for the mode of $p(y_{k:k+l} \, | \, x_k)$. For our numerical example, the approximation for the mode of $p(y_{k:k+l} \, | \, x_k)$ is obtained using an extended Rauch-Tung-Striebel (RTS) smoother \cite{sarkka2013bayesian}, initialized from a Gaussian distribution over $x_k$, with mean and covariance given by the empirical mean and covariance of $\{c_{k-1}(\xi_{k-1}^i) \}_{i=1}^n$.
\end{itemize}
We consider all the above with multinomial resampling, and also some of them with instead systematic resampling, the latter being indicated below by a suffix `sys'.

The performance of the particle filters in likelihood estimation is measured by computing
\begin{equation}
\text{Var}(\log Z_t) = \frac{1}{\tau} \sum_{j=1}^\tau (\log Z_t^j - \log \bar{Z}_t)^2,
\end{equation}
where $\tau$ is the number of samples and $\bar{Z}_t$ is the sample mean of $\{Z_t^j\}_{j=1}^\tau$. Our interest in this quantity is that the variability of $Z_t$ affects mixing when the particle filter is used within PMCMC. Generally speaking, higher variability degrades mixing. Probability computations are done with logarithms to avoid numerical problems.

The quality of the chain $\{\theta^j \}_{j=1}^\tau$ generated by the PMCMC algorithm can be assessed through the sample autocorrelation. Typically $\theta$ is a vector of parameters, say of length $p$, and the autocorrelation is computed for each 1-dimensional component $\theta_i$, $1 \leq i \leq p$, as
\begin{equation}
\text{ac}_i(l) = \frac{1}{\text{ac}_i(0)}\frac{1}{\tau-1} \sum_{j=1}^{\tau-l} (\theta^{j}_i-\bar{\theta}_i) (\theta^{j+l}_i - \bar{\theta}_i),
\end{equation}
where $l$ is the lag, $\tau$ is the number of samples in the chain, and $\bar{\theta}_i$ is the sample mean of $\{\theta_i^j\}_{j=1}^\tau$.
Since correlations in the MCMC chain contribute to the variance of the parameter estimate, we would like to see the autocorrelation approach zero rapidly for a good quality MCMC chain.

The autocorrelation can be used to compute a single summary number for the quality of the MCMC chain, called the effective sample size \cite{kass1998markov},  and given by
\begin{equation}
\tau_{\text{eff}} = \frac{\tau}{1+ 2 \sum_{l=1}^{\infty} \text{ac}(l)}.
\end{equation}
The effective sample size gives an approximation for the equivalent number of independent samples contained in the MCMC chain.

All the particle filters we tested were implemented in \textsc{Matlab} (R2014a).
Computations were performed using a MacBook Pro with 3\,GHz Intel i7 and 8\,Gb of memory.

\renewcommand{\figurename}{Fig. }
\setcounter{figure}{0}

\subsection{Positioning using range and bearing measurements}
\label{sec:ex1}

The first example we consider is a target tracking problem with nonlinear measurements, where the goal is to estimate the trajectory of a moving object e.g. a vehicle or a person using range and bearing measurements from a single measurement station. This is a prototypical problem in the literature on particle filters for target tracking, see e.g., \cite{ristic2004beyond, sarkka2013bayesian}.

The state $X=(R, V)$ consists of position $R=(R_1, R_2) \in \mathbb{R}^2$ and velocity $V=(V_1, V_2) \in \mathbb{R}^2$.
The dynamical model, formed by discretizing the constant velocity continuous-time stochastic model, is linear and given by
\begin{align}
X_{k+1} = \left[ \begin{array}{cc} \mathbf{I} & \mathbf{I} \Delta t \\ \boldsymbol{0} & \mathbf{I} \end{array} \right] X_k + \omega_k,
\end{align}
where $\omega_k$ is zero-mean Gaussian white noise with covariance
\begin{equation}
\mathbf{Q} = q^2 \left[ \begin{array}{cc} \Delta t^3/3 \mathbf{I} & \Delta t^2 /2 \mathbf{I} \\ \Delta t^2/2 \mathbf{I} & \Delta t  \mathbf{I} \end{array} \right]
\end{equation}
and $\Delta t$ is the time step between states.
The initial state is taken to be Gaussian with mean $\nu_0 = [100,100,0,0]^T$ and covariance chosen to reflect a relatively large uncertainty in the initial position,
\[
\mathbf{P}_0 = \left[ \begin{array}{cccc}10^2 & 0 & 0 & 0 \\ 0 & 10^2 & 0 & 0 \\ 0 & 0 & 10^{-3} & 0  \\ 0 & 0 & 0 & 10^{-3} \end{array} \right].
\]

The measurements are the range and bearing measured from a stationary measurement station located at coordinates $(0,0)$.
The measurements are modeled by
\begin{equation}
Y_k = h(X_k) + \zeta_k
\end{equation}
where
\begin{equation}
h(r,v) = \left[ \begin{array}{c} \| r \| \\ \arctan (r_2/r_1) \end{array} \right],
\end{equation}
and $\zeta_k$ is zero-mean Gaussian white noise, independent of $\omega_k$, with covariance
\[
\mathbf{R}=  \left[ \begin{array}{cc} \sigma^2_1 & 0 \\ 0 & \sigma^2_2  \end{array} \right].
\]

The unknown parameters are the process noise variance parameter $q^2$ and the measurement noise variances $\sigma^2_1$ and $\sigma^2_2$.
For the unknown parameters, we use independent inverse Gamma priors $\mathcal{IG}(a,b)$ with shape $a$ and scale $b$ parameters set to $a=b=0.1$ for measurement noises parameters, and to $a=1$ and $b=0.01$ for the process noise $q^2$.

\begin{figure}[!t]
\centering
\includegraphics[scale=0.25]{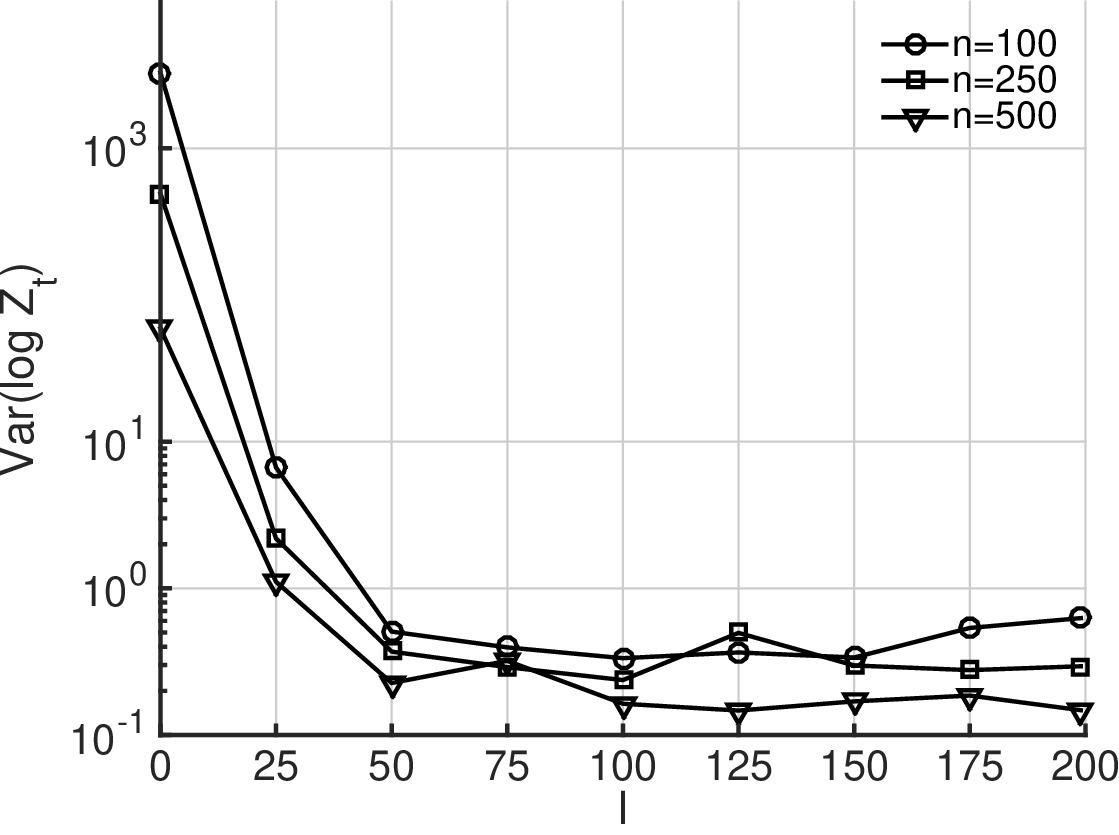}
\caption{Example A: Variance of $\log Z_{t}$ versus the parameter $l$ for twisted-BSPF-mode-sys with different number of particles. Results are averaged over 10 datasets and 30 simulations for each dataset. }
\label{fig:ex1_twisted_l}
\end{figure}

To test the performance of the different methods, we generated 10 datasets each consisting of $t=200$ measurements.
Fig. \ref{fig:ex1_twisted_l} shows the variance of $\log Z_t$ for the twisted-BSPF-mode-sys with different values of the twisting function parameter $l$.
It can be seen that increasing the value over $l=50$ does not give significant reduction in the variance.
In the subsequent tests we fix $l=50$.

\begin{figure}[!t]
\centering
\includegraphics[scale=0.23]{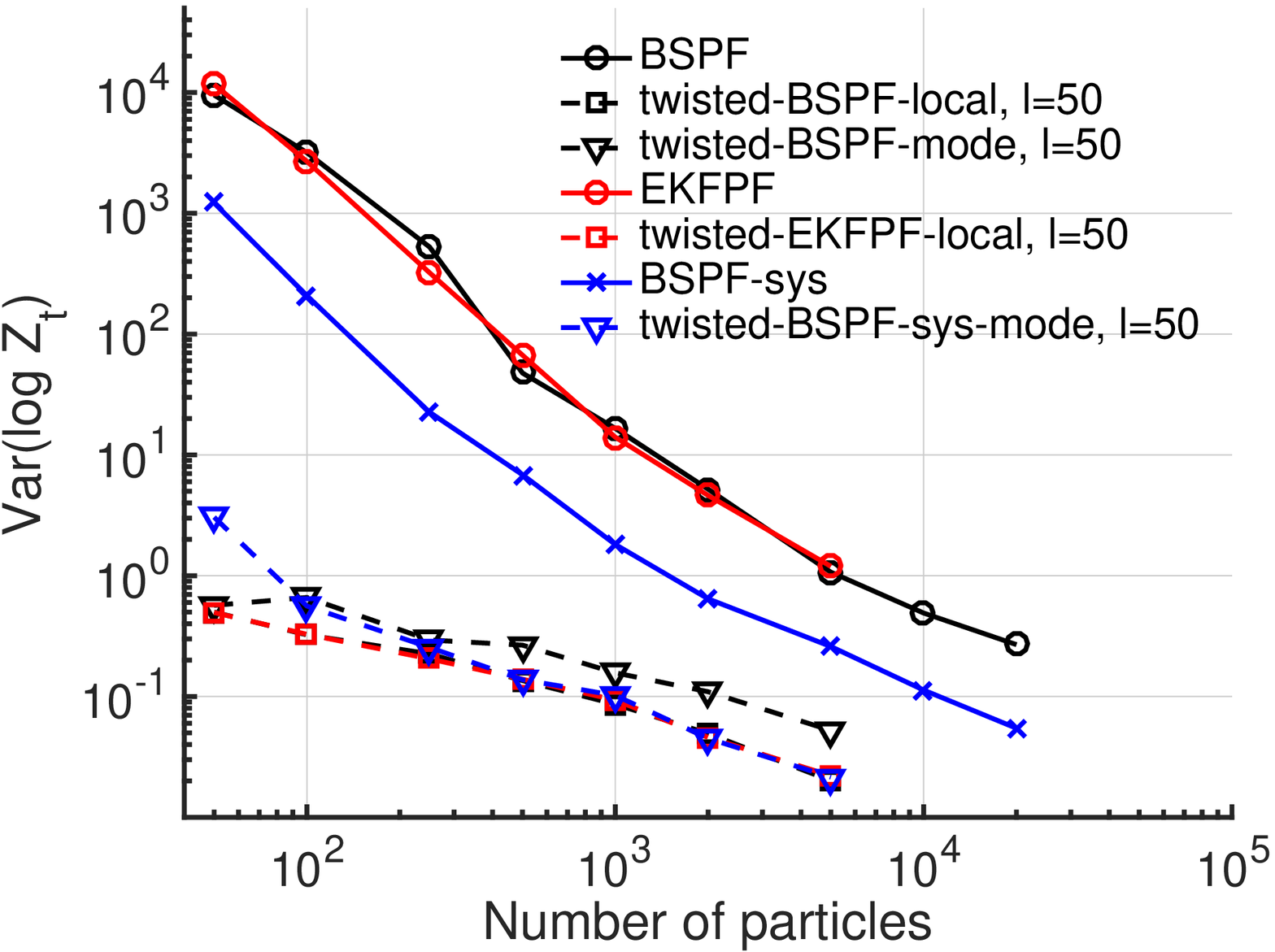} \hspace{0.05cm}
\includegraphics[scale=0.23]{ex1_varZ_time.eps} \\
\caption{Example A: Variance of $\log Z_{t}$ versus the number of particles $n$ (left) and time (right) for different particle filters. Parameters are fixed to the ground truth values. Results are averaged over 10 datasets and 30 simulations for each dataset. }
\label{fig:ex1_varZ}
\end{figure}

Fig. \ref{fig:ex1_varZ} shows the variance of $\log Z_t$ for the different methods as a function of the number of particles and computation time.
The twisted particle filters clearly outperform the non-twisted particle filters when looking at the $\log Z_t$ variance as a function of the number of particles.
However, the local linearization based twisted particle filters have a high computation time in this example.
Note also that since the EKF approximations for the importance distribution in EKFPF can be computed as a part of the local linearization for the twisting function, the computation times for twisted-BSPF-local and twisted-EKFPF-local are about the same.
The twisted-BSPF-mode algorithm, based on linearizing around the mode of the twisting function, is computationally much lighter and gives the lowest variance for the $\log Z_t$ in a given computation time.
For both the twisted and non-twisted particle filters using systematic resampling improves the results compared to the results using multinomial resampling.

We next analyze performance of the methods for generating samples using the Metropolis-Hastings PMCMC sampler.
Based on the results in Fig. \ref{fig:ex1_varZ} we chose the twisted-BSPF-mode-sys and BSPF-sys as the test methods.
We randomly chose one of the datasets as a test set and generated 20\,000 samples using the PMMH sampler.
An initial test run using the BSPF-sys with $n=5000$ particles was used to tune the proposal covariance, which was then held constant for the subsequent test runs.

\begin{figure}[!t]
\includegraphics[scale=0.23]{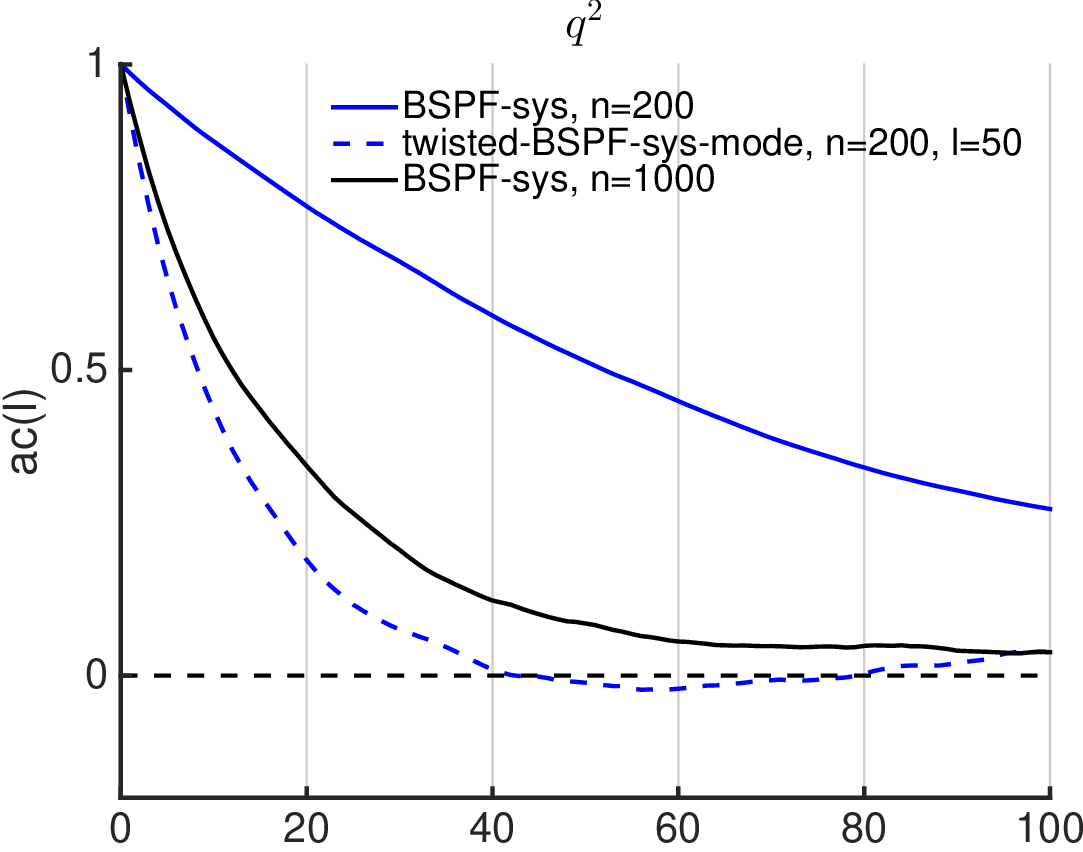} \hspace{0.05cm} \vspace{0.5cm}
\includegraphics[scale=0.23]{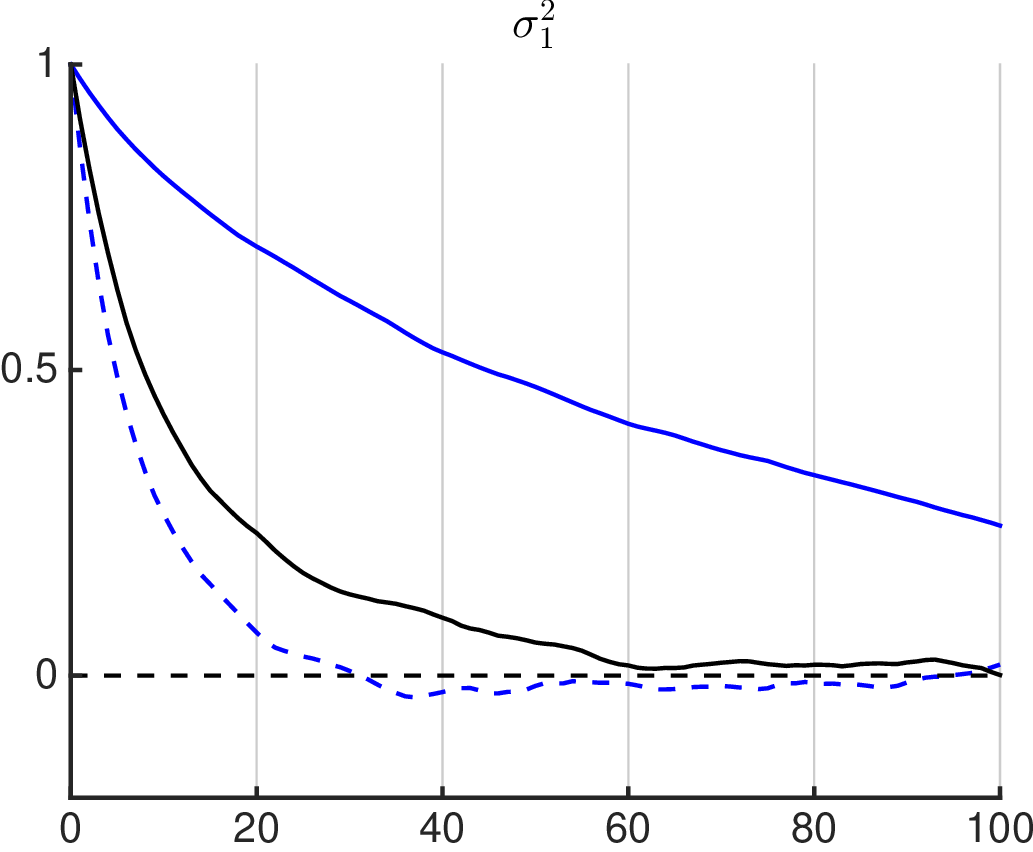}  \\
\includegraphics[scale=0.23]{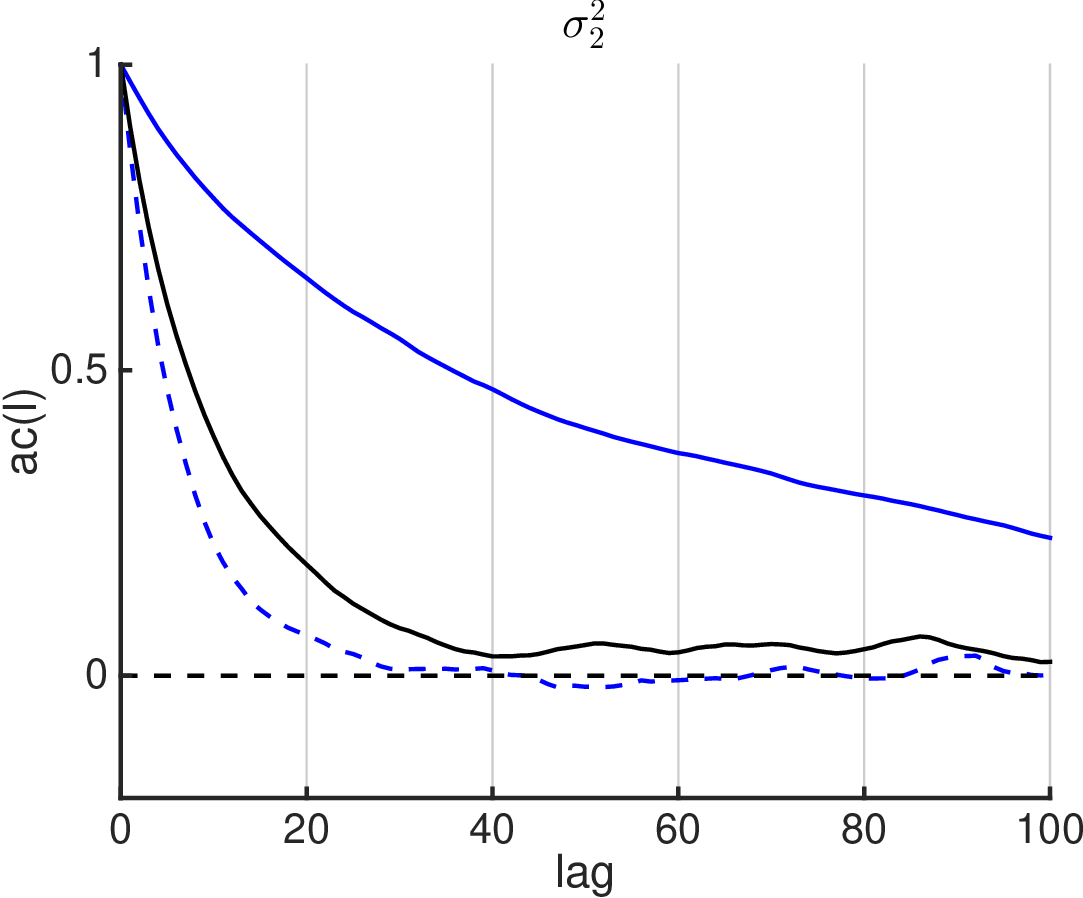}
\caption{Example A: Autocorrelation plots from a MCMC chain with 20\,000 samples generated using PMMH with BSPF and twisted-BSPF-mode. The autocorrelations are computed with burn in of 2500 samples excluded from the computations. }
\label{fig:ex1_autocorr}
\end{figure}

Fig. \ref{fig:ex1_autocorr} shows the autocorrelation performance for the methods.
The BSPF-sys with $n=1000$ particles has about the same computation time as twisted-BSPF-mode-sys with $n=250$ particles and $l=50$.
The twisted-BSPF-mode-sys has clearly better autocorrelation performance than the BSPF-sys with similar computation time.
The effective sample sizes and relative computation times are shown in Table \ref{table:ex1_Meff}. The relative computation time is obtained as the ratio of running time for each algorithm setting to that of twisted-BSPF-sys-mode with $50$ particles.
Generating 20\,000 samples with the twisted-BSPF-mode-sys with 50 particles took approximately 9 minutes.
The twisted-BSPF-mode-sys gives clearly larger effective sample sizes in less computational time than the BSPF-sys.

\begin{table}[!t]
\caption{Example A: The average effective sample sizes and relative computation times for the different particle filters. }
\label{table:ex1_Meff}
\center
\small{
\begin{tabular}{r r | l l}
Particle filter & $n$ & $\text{avg.}\;\tau_{\text{eff}}$ & rel. time  \\
\hline
BSPF-sys & 200 & 167.8 & 0.5 \\
& 500 & 404.6 & 1.0 \\
& 1000 & 540.6 & 1.7 \\
& 2000 & 753.5 & 3.4 \\
\hline
twisted-BSPF-mode-sys & 50 & 793.8 & 1.0 \\
& 100 & 890.0 & 1.2 \\
& 250 & 969.7 & 1.5 \\
& 500 & 1019.5 & 2.7
\end{tabular}
}
\end{table}

\begin{figure}[!t]
\centering
\includegraphics[scale=0.23]{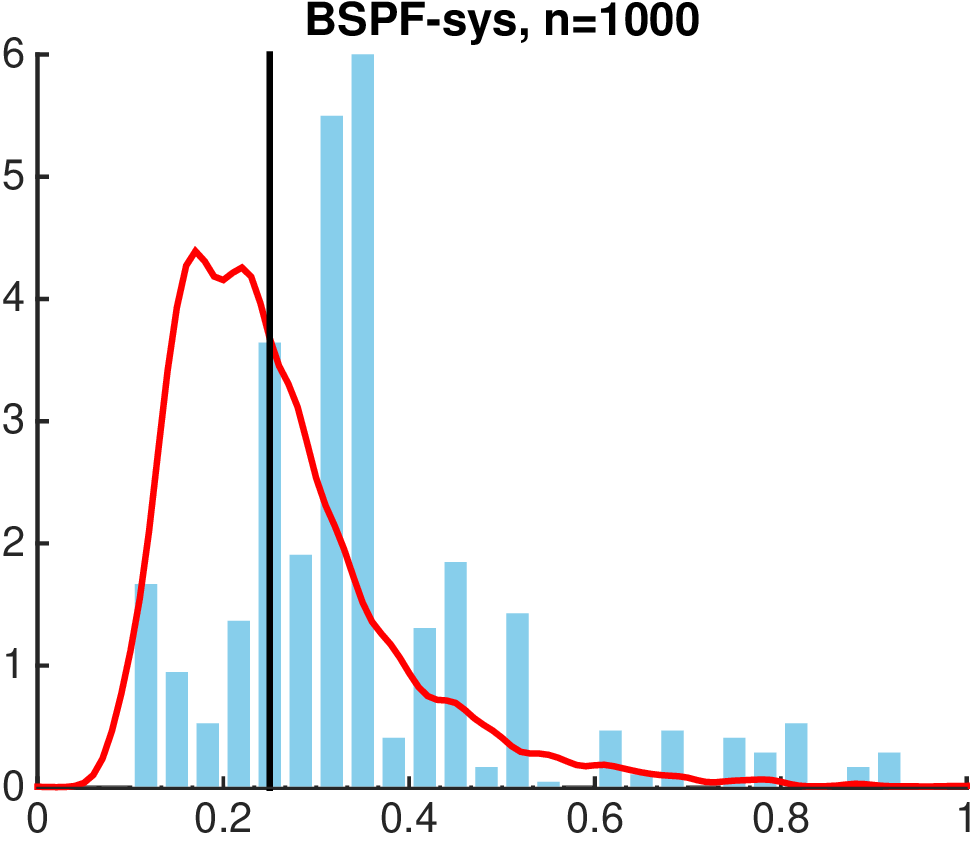} \hspace{0.05cm}
\includegraphics[scale=0.23]{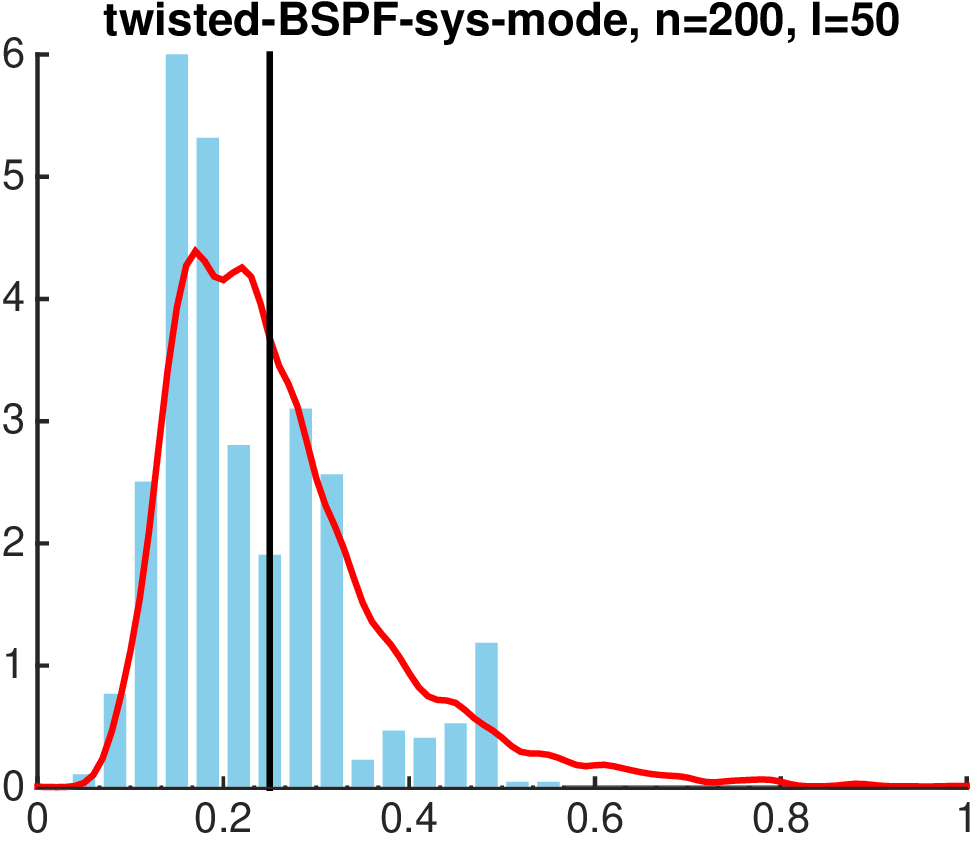} \\
\vspace{0.5cm}
\includegraphics[scale=0.23]{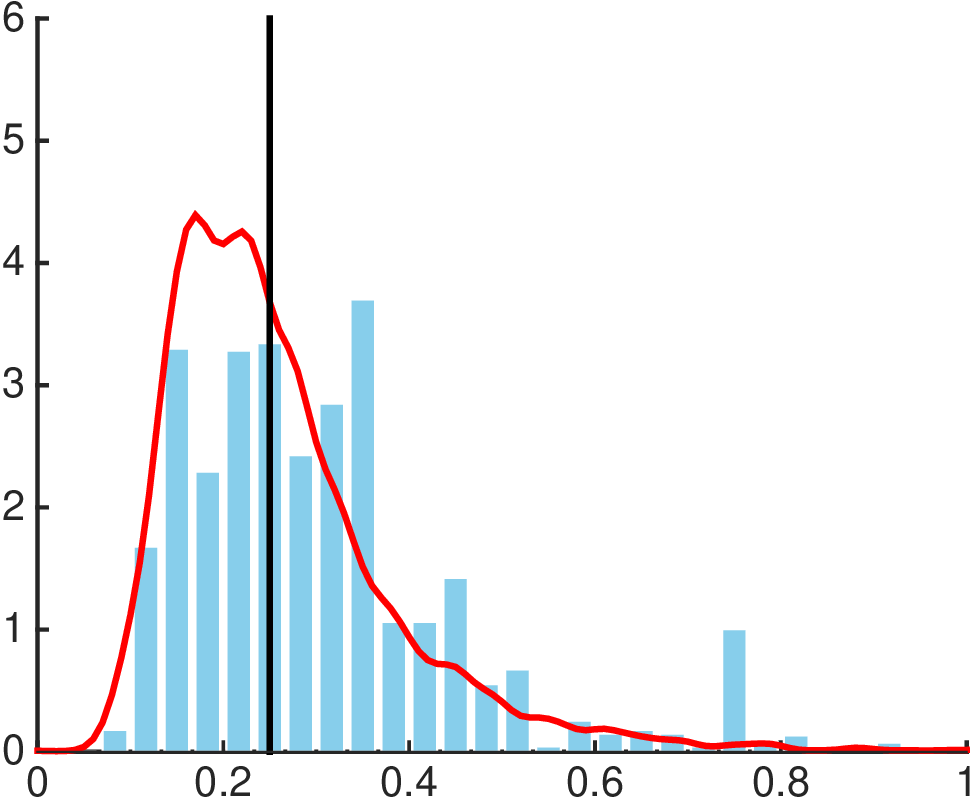} \hspace{0.05cm}
\includegraphics[scale=0.23]{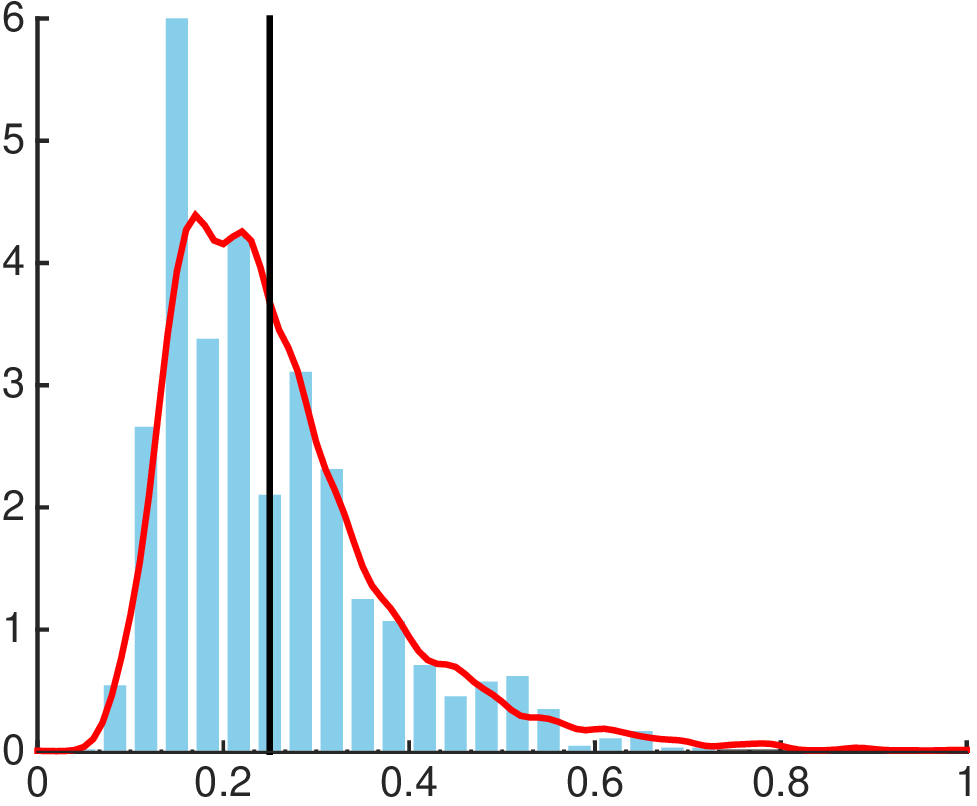} \\
\caption{Example A: Normalized histograms for parameter $q^2$ computed from PMCMC chains using BSPF-sys with $n=1000$ (left) and twisted-BSPF-mode-sys with $n=200$ and $l=50$ (right). Number of PMCMC samples used are 500 (top row) and 2000 (bottom row). The estimated true posterior (red line) is fitted to a separate PMCMC chain with 20\,000 samples. The ground truth value of the parameter is shown with a black line. }
\label{fig:ex1_hist}
\end{figure}

The convergence of the MCMC sequence is demonstrated in Fig. \ref{fig:ex1_hist} using normalized histograms computed from the MCMC chains.
The better mixing of the MCMC chain computed using twisted-BSPF-mode-sys is especially evident in the top row histograms computed using only a small number of samples.

A simple demonstration of the tracking performance using the estimated parameter values is shown in Table \ref{table:ex1_tracking}.
We used an EKF algorithm with parameters fixed to mean values of the MCMC chains with 500 samples.
The consistency value gives the fraction of times the true position is inside the 95\% confidence ellipsoid, averaged over all the time steps.
The 95\% confidence ellipsoid at time $k$ is given by
\[
(\mu_k - x_{\text{true}})^T \mathbf{\Sigma}_k^{-1}(\mu_k - x_{\text{true}}) =  F_{\chi_2^2}^{-1}(0.95),
\]
where $F^{-1}_{\chi_2^2}(0.95)$ is the value of the $\chi^2$ inverse cumulative distribution function with 2 degrees of freedom evaluated at $0.95$.
\begin{table}[!t]
\caption{Example A: Tracking performance using EKF}
\label{table:ex1_tracking}
\center
\small{
\begin{tabular}{r | l l }
PMCMC method & RMSE & 95\% cons. \\
\hline
BSPF-sys, $n=1000$ & 13.2 & 0.92 \\
twisted-BSPF-sys-mode, $n=200$, $l=50$ & 12.4 & 0.91
\end{tabular}
}
\end{table}

\subsection{Positioning using RSS measurements}

As a second example, we consider estimating the parameters of a received signal strength (RSS) measurement model in an indoor positioning scenario using Bluetooth measurements.
As the user moves inside the building, the positioning device measures the attenuated signal from Bluetooth base stations situated at known locations.
Given a suitable model for the signal attenuation, the measurements give information about the distance between the positioning device and the base stations.
Combined with a motion model, we can then use the measurements to track the user's movements inside the building.

For this example, we use a simple two parameter empirical model for the signal attenuation \cite{hatay1980empirical}.
The base station specific parameters, together with any other unknown parameters (e.g. noise variances), are estimated using a learning dataset.
We consider a full Bayesian approach and use the PMCMC algorithm to draw samples from the true parameter posterior distributions.
The samples can then be used to compute point estimates or integrate out the parameters in subsequent positioning phases.

\begin{figure}[!t]
\centering
\includegraphics[scale=0.34]{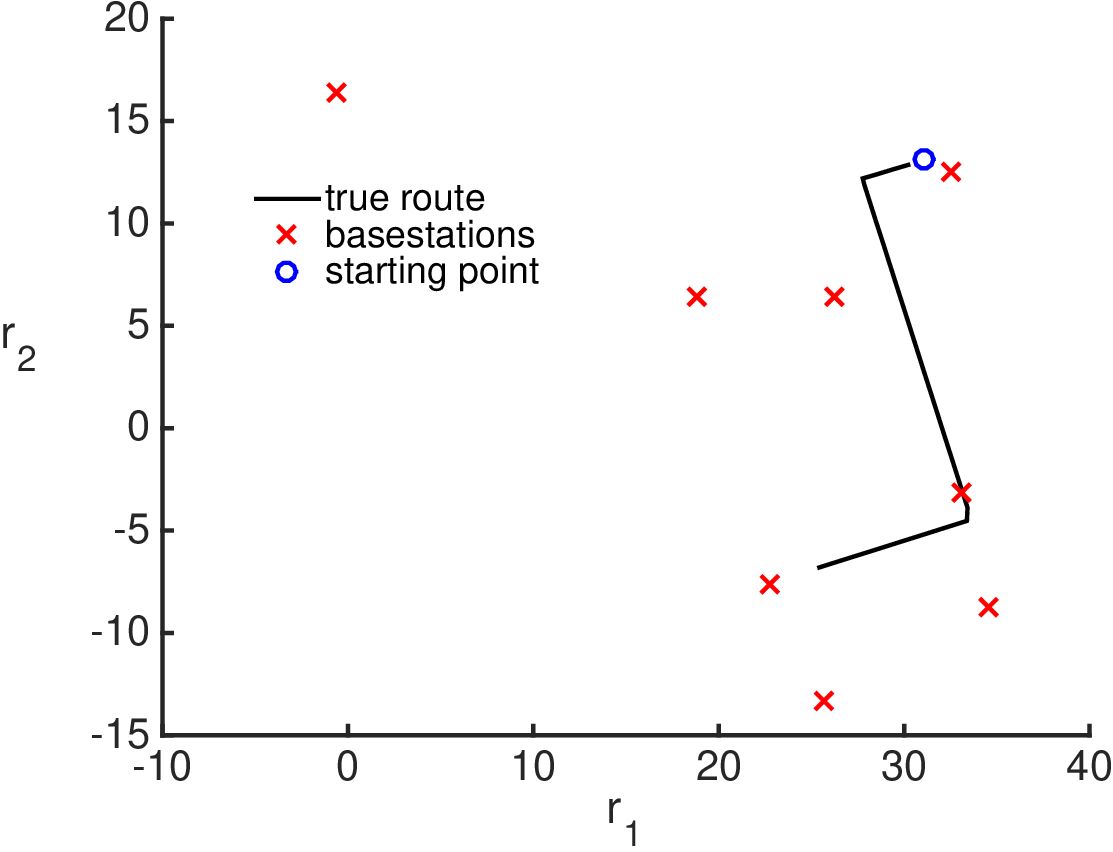}
\caption{Example B: The indoor positioning scenario.}
\label{fig:ex2_data}
\end{figure}

We use a real data set collected in a building at the Tampere University of Technology.
This data consists of RSS measurements from 8 different base stations with a total of $t=54$ RSS measurement vectors.
The locations of the base stations and the true route is shown in Fig. \ref{fig:ex2_data}; the true route was obtained by having the user manually indicate his location on a map at regular intervals.
The number of elements in the RSS measurement vector at a single time point ranges from 1 to 7, with an average number of about 5 elements per time point.

The dynamical model is the same as in the first example in Section \ref{sec:ex1}.
The initial state is taken to be Gaussian with mean $\nu_0$ and covariance $\mathbf{P}_0$.
For this example, we fix the position components of the initial mean to the true location, and set the velocity components to zero.
The initial covariance is the same as in the first example.

The measurements are modelled as
\[
Y_k = h_k(X_k) + \zeta_k,
\]
where $h_k(x)$ is a vector with elements $h^i(x)$, $i\in \mathcal{I}_k$, where $\mathcal{I}_k$ contains the indices of the base stations whose RSS are measured at time $k$,  $\zeta_k$ is a zero-mean Gaussian vector, independent of $\omega_k$, with covariance $\mathbf{R}= \sigma^2 \mathbf{I}$,
and the RSS measurement function is \cite{hatay1980empirical}
\[
h^i(r,v) = \rho_i - 10 \lambda_i \log_{10} \| r_{\text{BS},i} - r \|, \quad 1 \leq i \leq n_{\text{BS}},
\]
where $r_{\text{BS},i}$ are the locations of measurement stations, $\lambda_i$ and $\rho_i$ are the base station specific parameters, and $n_{\text{BS}}$ is the number of base stations.

\begin{figure}[!t]
\centering
\includegraphics[scale=0.23]{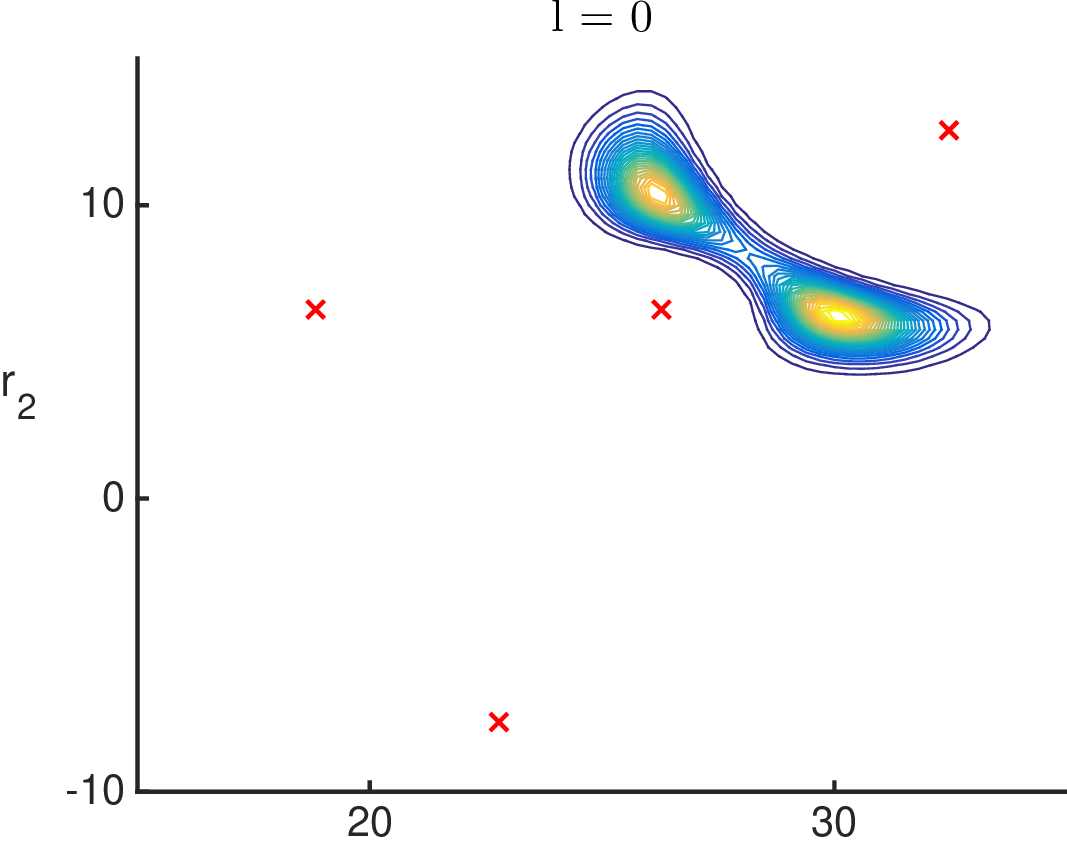} \hspace{0.05cm}
\includegraphics[scale=0.23]{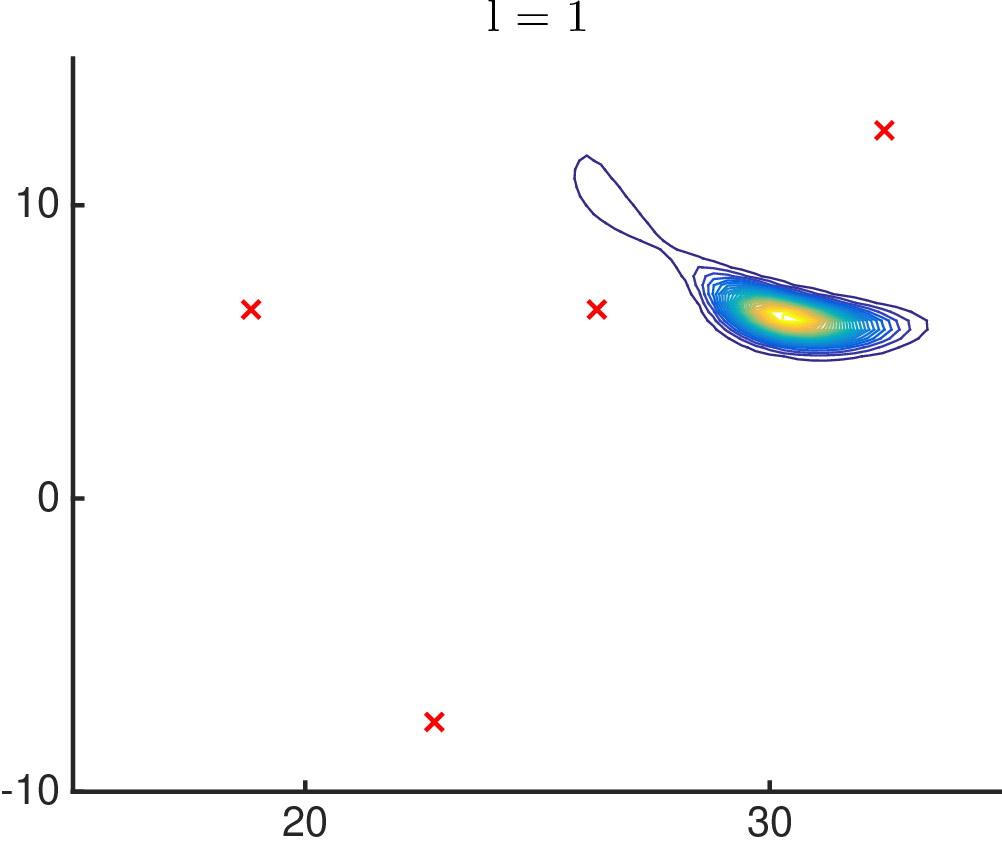}  \\
\vspace{0.5cm}
\includegraphics[scale=0.23]{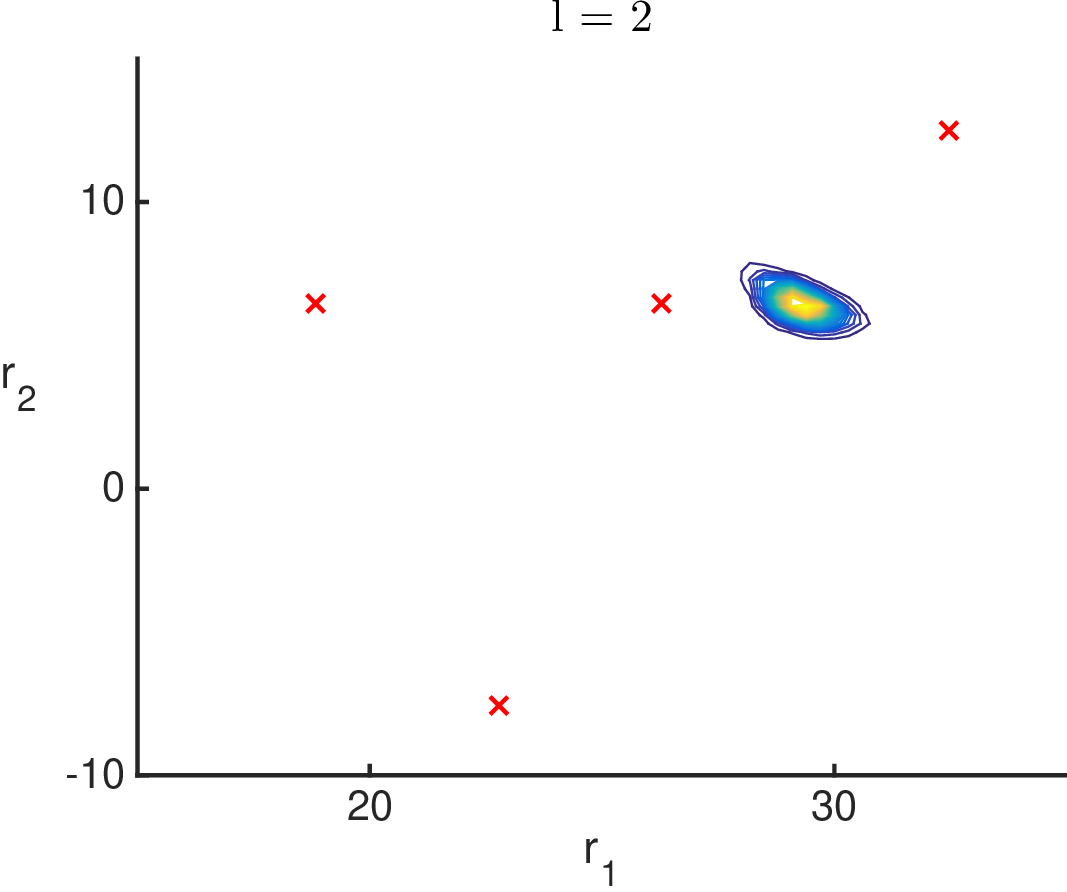} \hspace{0.05cm}
\includegraphics[scale=0.23]{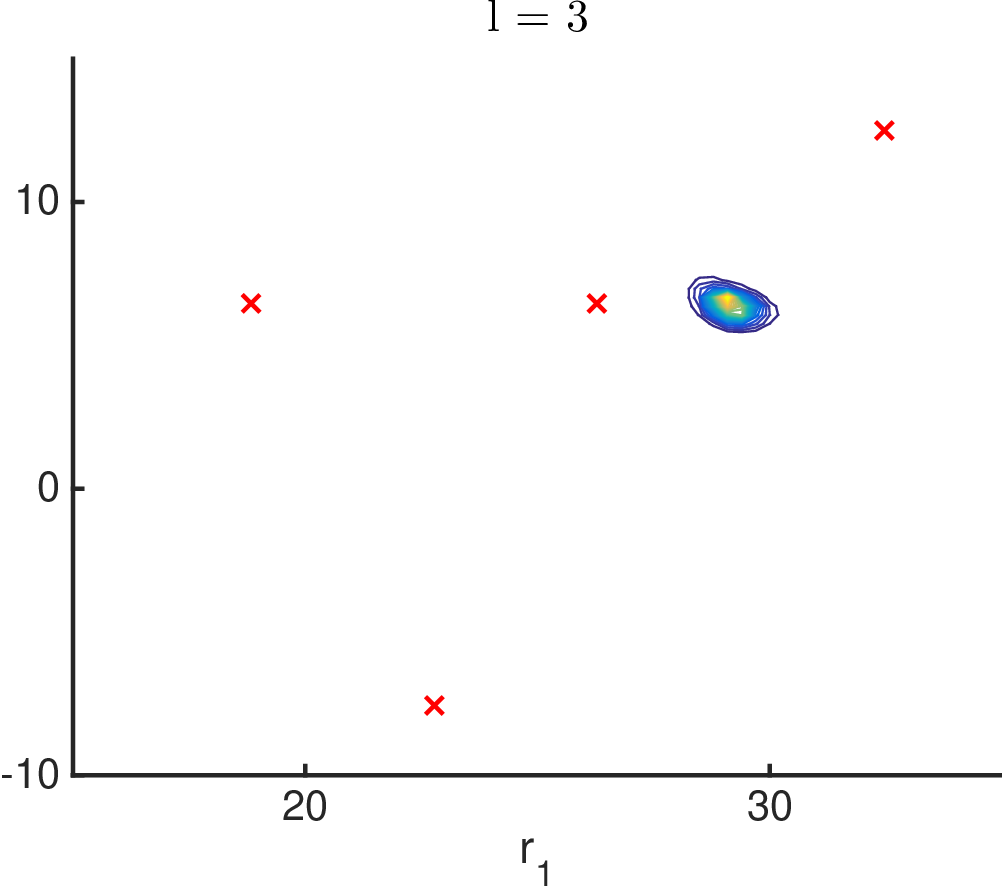}
\caption{Example B: Illustration of the behaviour of $p(y_{k:k+l} \, | \, x_k)$ as the number of measurements $l$ increases. The values for $l > 0$ are computed by running a particle filter separately for each point on a dense grid for the position $r_k$. The velocity $v_k$ is kept fixed for this example plot.}
\label{fig:ex2_likelihood}
\end{figure}

The measurement likelihood is strongly non-Gaussian and can be multimodal, depending on the geometry of the base station locations.
However, the term $p(y_{k:k+l} \, | \, x_k)$ becomes concentrated on a single mode as the number of measurements $l$ increases (see Fig. \ref{fig:ex2_likelihood}).
This allows us to reduce the computation time of the twisted particle filter by using the linearization around the mode of $p(y_{k:k+l} \, | \, x_k)$, described in Section \ref{sec:mode_linearization}, when $l$ is sufficiently large.

The unknown parameters are the transition noise variance parameter $q^2$, the process noise variance $\sigma^2$, and the measurement model parameters $\lambda_i$ and $\rho_i$, $i=1,\ldots, n_{\text{BS}}$. Priors for the parameters are chosen as follows.
For the noise variance parameters, we use independent inverse Gamma priors $\mathcal{IG}(a,b)$ with shape $a$ and scale $b$ parameters set to $a=b=0.1$ for measurement noise $\sigma^2$, and to $a=1$ and $b=0.01$ for the process noise $q^2$.
For the path-loss exponents $\lambda_i$, we use independent gamma priors $\mathcal{G}(a,b)$, with shape parameter $a=3.8$ and scale parameter $b=1.6$.
For the parameters $\rho_i$ we use independent Gaussian priors with zero mean and variance $70^2$.

We first determine an initial approximation for the posterior mean by generating 10\,000 samples using the PMMH and BSPF with $n=5000$ particles.
For this relatively high dimensional problem, we found that it was necessary to use a component-wise update, also called Metropolis-within-Gibbs, in the PMMH sampler.
The parameters are updated in $n_{\text{BS}}+1$ blocks of 2 variables, with the blocks consisting of $(\lambda_i, \rho_i)$, for $i=1,\ldots,n_{\text{BS}}$ and $(q^2, \sigma^2)$ for the final block.
For each block, we have an independent Gaussian random walk proposal, with covariance tuned during the initial PMMH run and kept fixed in the subsequent test runs.

\begin{figure}[!t]
\centering
\includegraphics[scale=0.25]{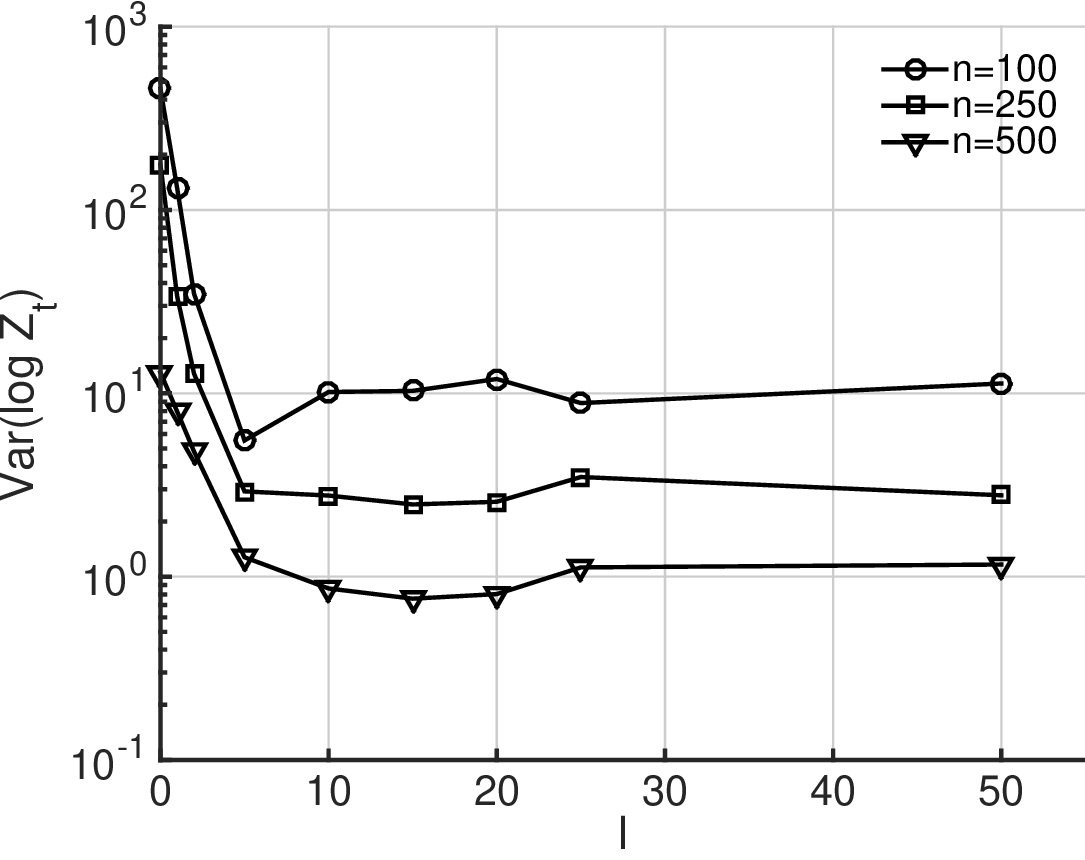}
\caption{Example B: Variance of $\log Z_{t}$ versus the parameter $l$ for twisted-BSPF-local using multinomial resampling with different number of particles. Results are computed from 100 simulations. }
\label{fig:ex2_twisted_l}
\end{figure}

Fig. \ref{fig:ex2_twisted_l} shows the variance of $\log Z_{t}$ for the twisted-BSPF-local as a function of the parameter $l$, with the unknown parameters fixed to the mean values from the initial test run.
It can be seen that increasing $l$ over 20 does not generally improve the results and can lead to larger variance of the estimate.
This is most likely caused by the gradually increasing linearization errors in the computation of the twisting function using the EKF, meaning that for large $l$ we have a slightly poorer approximation of the optimal twisting function.
For the following tests, we use a fixed $l=10$ for all the tested twisted particle filters.

\begin{figure}[!t]
\centering
\includegraphics[scale=0.23]{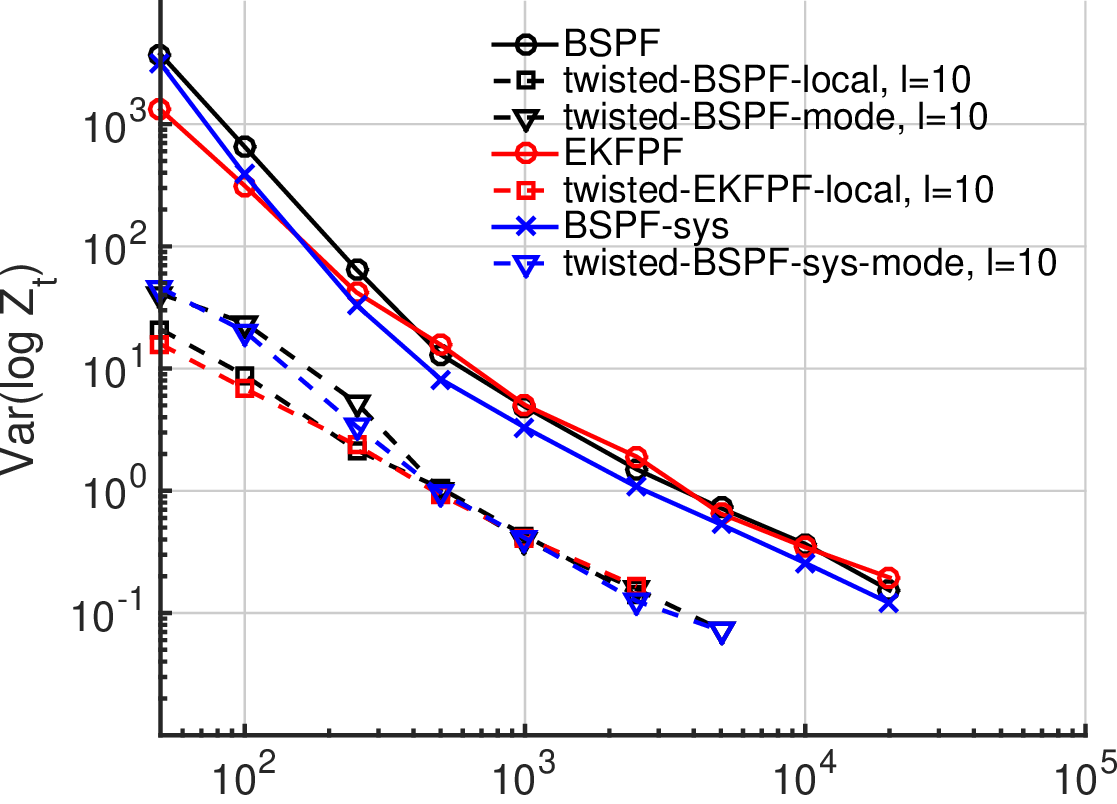} \hspace{0.05cm}
\includegraphics[scale=0.23]{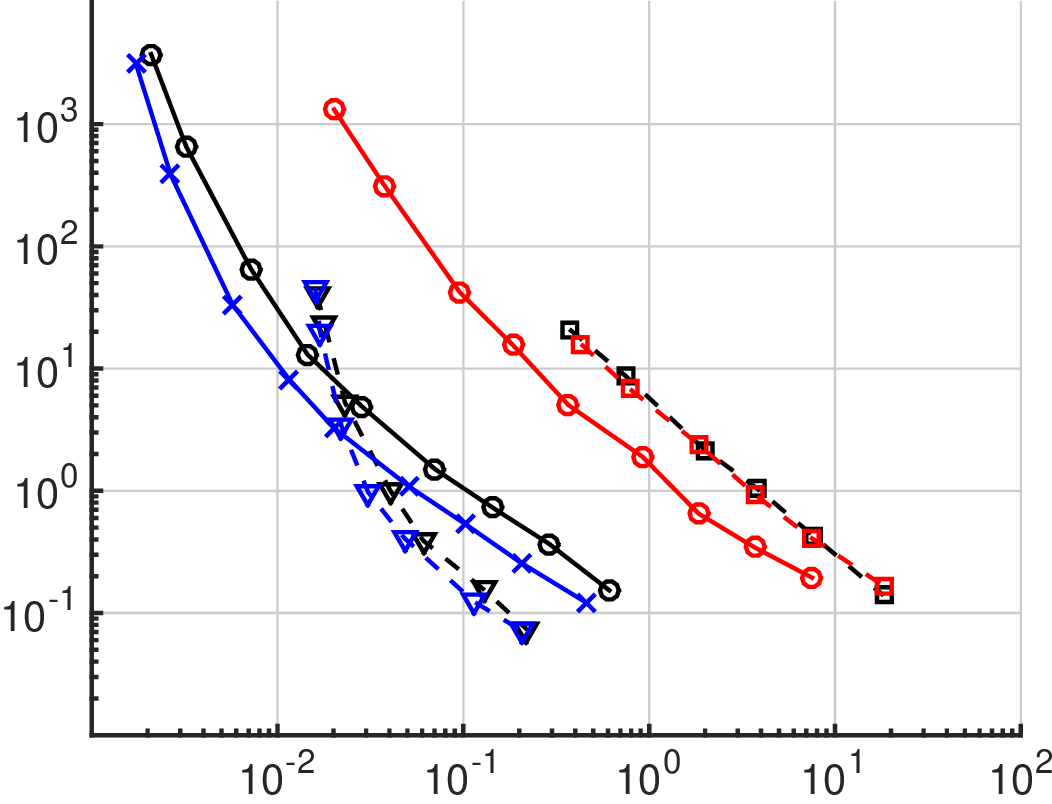} \\
\vspace{0.5cm}
\includegraphics[scale=0.23]{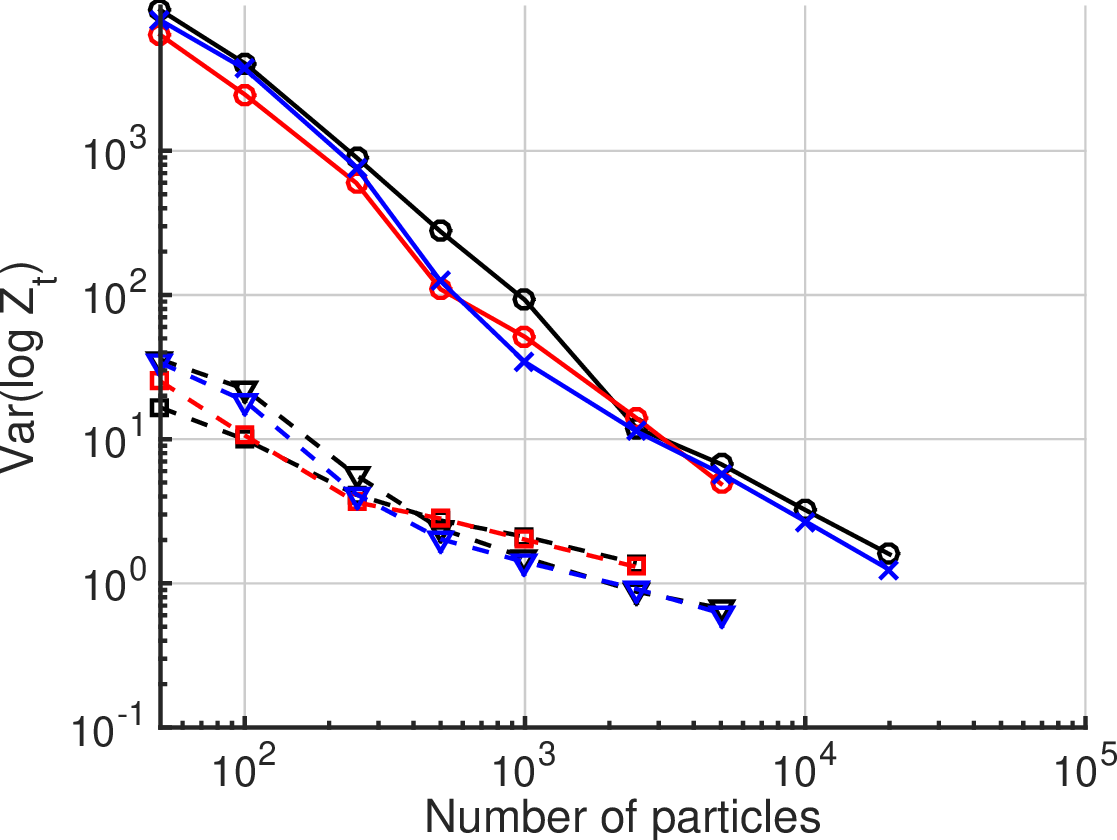} \hspace{0.05cm}
\includegraphics[scale=0.23]{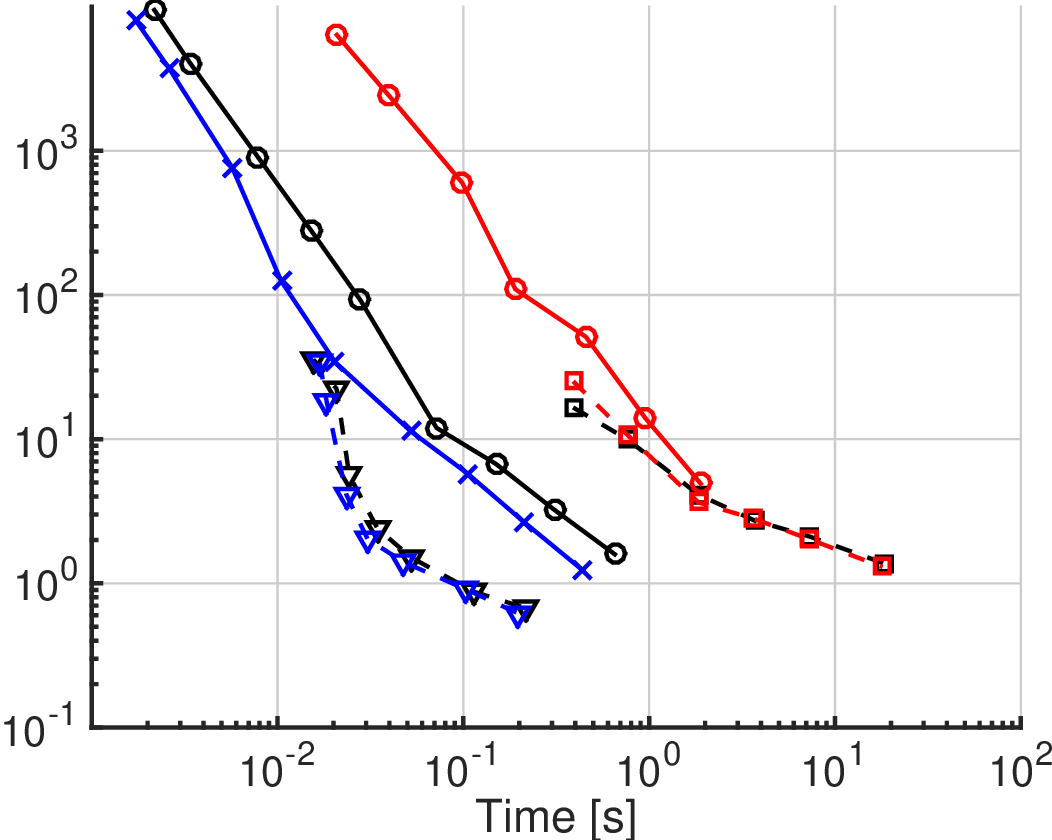}
\caption{Example B: Variance of $\log Z_{t}$ versus the number of particles $n$ (left) and time (right) for different particle filters. Parameters are fixed to a posterior mean estimate (upper row) and to a random value chosen from test PMCMC chain (lower row). Results are computed from 500 simulations. }
\label{fig:ex2_varZ}
\end{figure}

Fig. \ref{fig:ex2_varZ} shows the variance of the different particle filters as a function of number of particles and computation time.
Parameters were first fixed to the posterior mean estimate from the initial test run and then to a value chosen from the initial PMCMC chain, to test how the particle filters perform for parameter values away from the mean.

The results are similar as in the first example.
All the tested twisted particle filters clearly outperform the non-twisted particle filters when looking at the number of particles needed for a specific log-likelihood variance.
In a given computation time, the twisted-BSPF-mode-sys gives the lowest variance for $\log Z_t$.
Using systematic resampling improves the results for both twisted and non-twisted particle filters.

\begin{figure}[!t]
\centering
\includegraphics[scale=0.23]{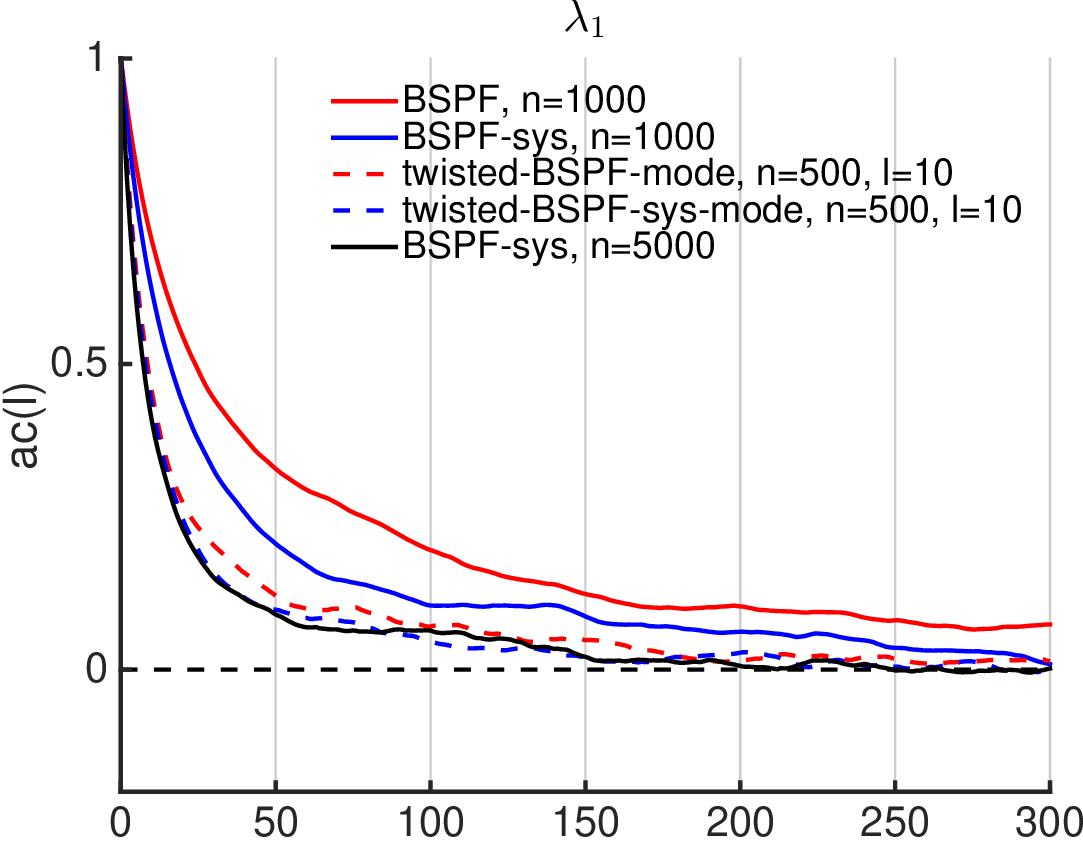} \hspace{0.05cm}
\includegraphics[scale=0.23]{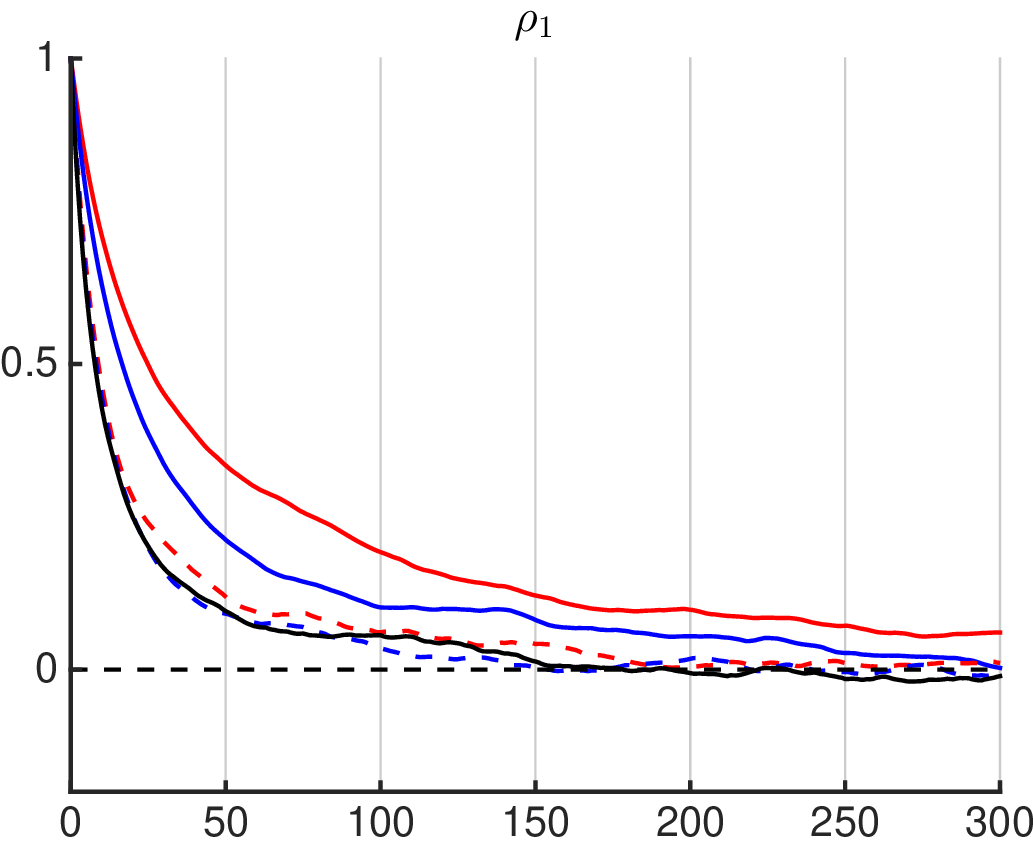} \\
\vspace{0.5cm}
\includegraphics[scale=0.23]{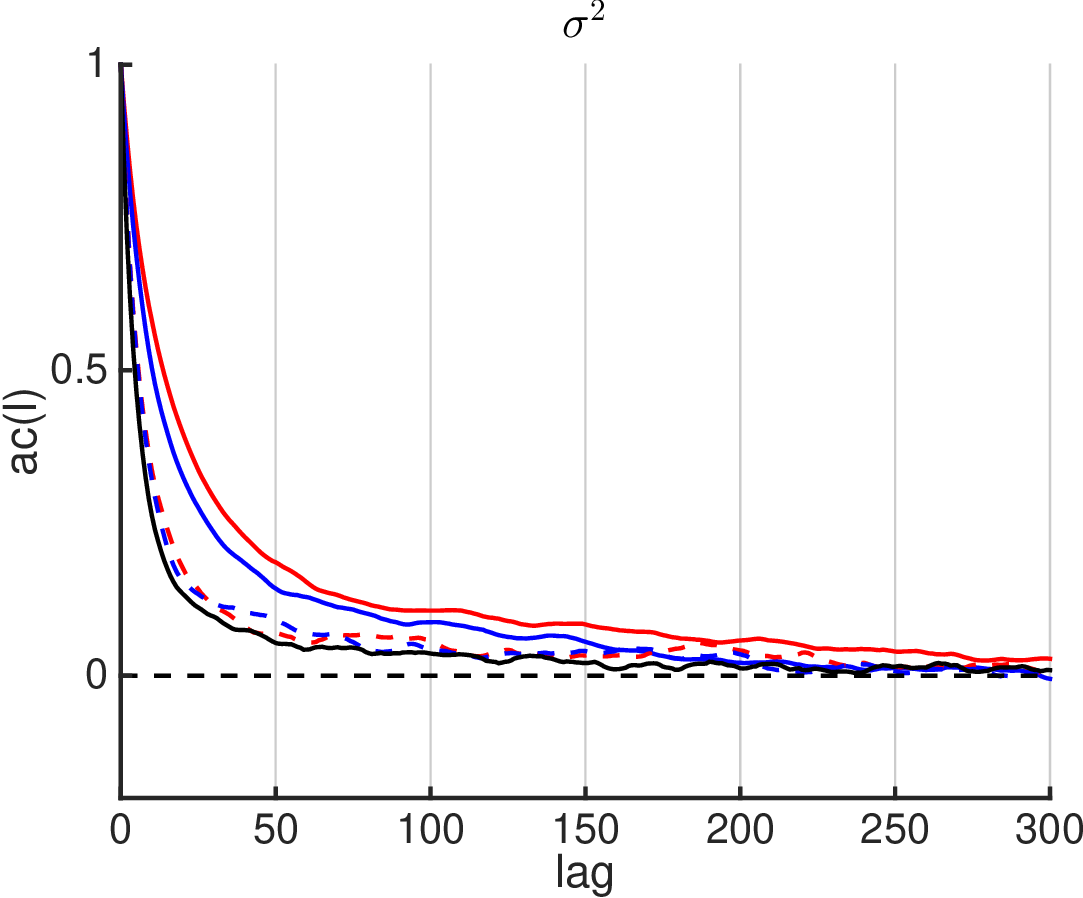} \hspace{0.05cm}
\includegraphics[scale=0.23]{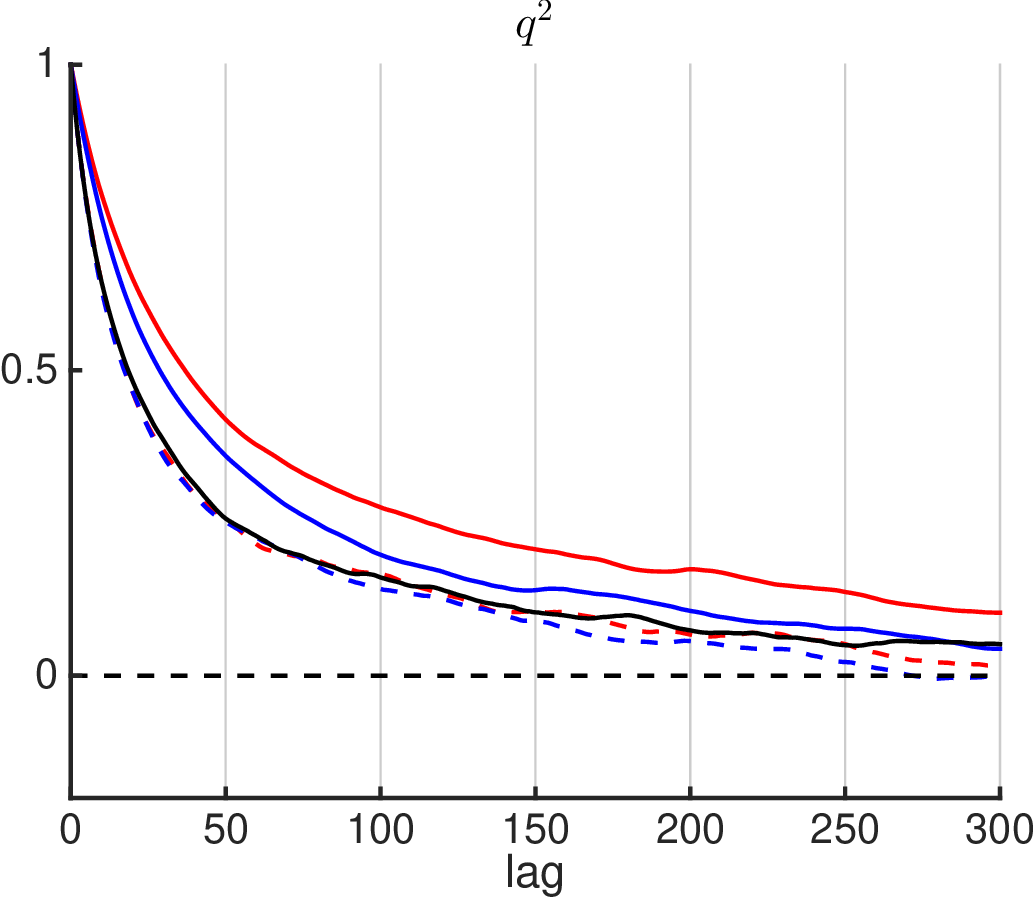}
\caption{Example B: Average autocorrelation plots from 10 MCMC chains generated using PMMH with BSPF and twisted-BSPF-mode. The plots are for parameters $\lambda_1$ (top left), $\rho_1$ (top right), $\sigma^2$ (bottom left) and $q^2$ (bottom right). The autocorrelations are computed from chains with \mbox{10 000} samples with burn in of 1000 samples excluded from the computations. }
\label{fig:ex2_autocorr}
\end{figure}

We proceed by comparing two of the most promising particle filters, i.e. the BSPF and twisted-BSPF-mode, in generating samples using the PMMH sampler.
For each particle filter, we generated a total of 100\,000 samples using 10 independent chains of 10\,000 samples.
Fig. \ref{fig:ex2_autocorr} shows the average autocorrelation plots over the 10 chains for base station parameters $\lambda_1$ and $\rho_1$, and noise variances $\sigma^2$ and $q^2$.
The BSPF-sys has clearly better performance compared to the BSPF with the same number of particles, as was expected from the log-likelihood variance results.
However, BSPF-sys still needs a significantly larger number of particles and longer computation times (see Table \ref{table:Meff}) to reach the same autocorrelation performance as twisted-BSPF-mode.

The average effective sample sizes over all the parameters and relative computation times are shown in Table \ref{table:Meff}. The relative computation time is obtained as the ratio of running time for each algorithm setting to the running time of the twisted-BSPF-mode algorithm with $250$ particles.
Generating 10\,000 samples with twisted-BSPF-mode with 250 particles took approximately 33 minutes.

 Results show that the two tested twisted particle filters give clearly the largest effective sample size with a given number of particles and in a given computational time.
For the twisted particle filters, the effect of using systematic resampling is relatively small, with the systematic resampling giving slightly better results especially for a large number of particles.

\begin{table}[!t]
\caption{Example B: The average effective sample sizes and relative computation times for the different particle filters. }
\label{table:Meff}
\center
\small{
\begin{tabular}{r r | l l}
Particle filter & $n$ & $\text{avg.}\;\tau_{\text{eff}}$ & rel. time  \\
\hline
BSPF & 1000 & 59.6 & 1.1 \\
& 2000 & 117.0 & 2.1 \\
& 5000 & 146.0 & 5.7 \\
& 10000 & 171.5 & 11.8 \\
\hline
BSPF-sys & 1000 & 81.1 & 0.9 \\
 & 2000 & 124.1 & 1.7 \\
 & 5000 & 165.5 & 4.5 \\
 & 10000 & 191.8 & 9.6 \\
\hline
twisted-BSPF-mode & 250 & 111.0 & 1.0 \\
 & 500 & 141.5 & 1.3 \\
 & 1000 & 162.3 & 2.1 \\
 & 2000 & 189.0 & 3.5 \\
\hline
twisted-BSPF-sys-mode & 250 & 110.0 & 1.0 \\
 & 500 & 149.6 & 1.3 \\
 & 1000 & 168.9 & 2.0 \\
 & 2000 & 199.9 & 3.4
\end{tabular}
}
\end{table}

A simple demonstration of the tracking performance using the estimated parameter values is shown in Table \ref{table:tracking}.
We used an EKF algorithm with parameters fixed to mean values of the respective MCMC chains, computed from 10\, 000 MCMC samples.
It should be noted that for our example, only a small amount of data was available, and for this reason the offline parameter estimation and online tracking were computed using the same data set.
In reality, one would use a separate, comprehensive data set for parameter estimation.

\begin{table}[!t]
\caption{Example B: Tracking performance using EKF}
\label{table:tracking}
\center
\small{
\begin{tabular}{r | l l }
PMCMC method & RMSE & 95\% cons. \\
\hline
BSPF-sys, $n=2000$ & 4.6 & 0.09 \\
twisted-BSPF-sys-mode, $n=500$, $l=10$ & 4.5 & 0.11
\end{tabular}
}
\end{table}

\section{Conclusion}\label{sec:conclusion}
Our numerical results indicate that twisted particle filters can give efficiency gains for marginal likelihood approximation and parameter estimation via PMCMC. The performance gains shown in Tables \ref{table:ex1_Meff} and \ref{table:Meff} illustrate a speed-up of about 3-5 times for the same average effective sample size, compared to standard methods. Of course, the amount of speed-up is implementation dependent, and in our implementations we have not gone to great lengths to optimize performance of the twisted particle filter, so larger gains may well be possible. On the other hand, the efficiency of the twisted particle filter rests on the choice of the twisting functions $\psi_k$, and the ability to choose a ``good'' $\psi_k$ is of course problem dependent.

For our purposes, a sufficient choice for $\psi_k$ was obtained by using an EKF based linearization of the non-linear model functions.
However, for problems where the EKF based methods fail to deliver a good approximation for optimal $\psi_k$, the presented algorithms could be modified to use linearization based on other types of Gaussian filters e.g. unscented Kalman filter or other sigma-point Gaussian filters described for example in \cite{sarkka2013bayesian}.

Further research should be conducted to determine the best approach for approximating the optimal twisting function in the case of multimodal $p(y_{k:k+l} \, | \, x_k)$. A possible solution could be to use mixture approximations with each component formed by linearizing the model functions around one of the modes.

There are also various other aspects of PMCMC methodology which could be developed around twisted particle filters, for example by deriving a PMMH algorithm to sample from $p(\theta,x_{0:t}|y_{0:t})$ rather than just $p(\theta|y_{0:t})$, and in deriving particle Gibbs samplers, along the lines of those introduced in \cite{andrieu2010particle}.

\appendices
\numberwithin{equation}{section}
\section{Proof of Proposition \ref{prop:unbiased}}
\label{app:proof_prop1}

Define functions $(\eta_k)_{k=0}^{t}$ recursively as $\eta_{t}(x_{t}) := 1$ and $\eta_{k-1}(x_{k-1}) := \int_{\mathbb{X}} g_k(y_k \, | \, x_k)f_k(x_k \, | \, x_{k-1}) \eta_k(x_k) \, dx_k$ for $t \geq k \geq 1$.
For any $1 \leq k \leq t$ we have
\begin{align}
& \mathbb{E}\left[ Z_k \frac{\sum_{i=1}^nW_k^i \eta_k(\xi_k^i)}{\sum_{i=1}^n W_k^i} \, \big| \, \mathscr{F}_{k-1}\right] \nonumber \\
 & = Z_{k-1} \mathbb{E}\left[ \frac{1}{n}\sum_{i=1}^n \int_{\mathbb{X}} W_k^i \eta_k(\xi_k^i)q_k(d\xi_k^i \, | \, \mathscr{L}_{k-1}^{r_{k-1}^i(U_{k-1})}) \, \big| \, \mathscr{F}_{k-1}\right] \nonumber \\
 & = Z_{k-1} \mathbb{E}\left[ \frac{1}{n}\sum_{i=1}^n  \eta_{k-1}(\xi_{k-1}^{r_{k-1}^i(U_{k-1})})\, \big| \, \mathscr{F}_{k-1}\right] \nonumber \\
 & =  Z_{k-1} \frac{\sum_{i=1}^nW_{k-1}^i \eta_{k-1}(\xi_{k-1}^i)}{\sum_{i=1}^n W_{k-1}^i}, \label{eq:Zk_rec}
\end{align}
where the second equality follows by plugging in $W_k^i$ and the final equality by using Assumption \ref{ass:resampling}.
We now have
\begin{align*}
\mathbb{E}[Z_t] & = \mathbb{E}\left[Z_0  \frac{\sum_{i=1}^nW_0^i \eta_0(\xi_0^i)}{\sum_{i=1}^nW_0^i} \right] \\
& =\frac{1}{n}\sum_{i=1}^n \mathbb{E}\left[W_0^i \eta_0(\xi_0^i)\right]= p(y_{0:t}),
\end{align*}
where the first equality follows by using (\ref{eq:Zk_rec}) repeatedly and the final equality by plugging in $W_0^i$ and taking the expectation.

\section{Proof of Theorem \ref{thm:twisted_unbiased}}
\label{app:proof_theorem1}

It was already established in Section \ref{sec:twisted_pf} that lines 1 to 5 in Algorithm \ref{alg:twisted_pf} draw $\xi_0$ from $\widetilde{\mathbf{M}}_0$.
We next show that lines 11 to 17 in Algorithm \ref{alg:twisted_pf} draw $\xi_k$ and $U_{k-1}$ from $\widetilde{\mathbf{M}}_k$.
Plugging in $\mathbf{M}_k$ to  (\ref{eq:particle_filter_tk}) we get
\begin{align}
& \widetilde{\mathbf{M}}_k(d \xi_k, d u_{k-1} \, | \, \mathscr{F}_{k-1}) \nonumber \\
 & \propto \sum_{s=1}^n \mathcal{U}(d u_{k-1}) q_k(d\xi_k^s \, | \, \mathscr{L}_{k-1}^{r_{k-1}^s(u_{k-1})}) \psi_{k}(\mathscr{L}_{k-1}^{r_{k-1}^s(u_{k-1})},\xi_k^s) \nonumber \\
 & \quad \quad \quad \quad \cdot \prod_{i \neq s} q_k(d\xi_k^i \, | \, \mathscr{L}_{k-1}^{r_{k-1}^i(u_{k-1})}).\label{eq:tildeMk}
\end{align}
We recognize this as a mixture form. So to sample $\xi_k$ and $U_{k-1}$, we first draw the mixture component $S_k$ on $\{1,\ldots,n\}$ with probabilities
\begin{align*}
& \widetilde{\mathcal{S}}_k(S_k = s) \propto \int_{[0,1]^m}\mathcal{U}(du_{k-1}) \nonumber \\
 & \cdot \int_{\mathbb{X}^n} q_k(d x_k^s \, | \, \mathcal{L}_{k-1}^{r_{k-1}^s(u_{k-1})} \psi_k(\mathcal{L}_{k-1}^{r_{k-1}^s(u_{k-1})}, x_k^s) \\
  & \quad \quad \cdot \prod_{i \neq s} q_k(dx_k^i \, | \, \mathscr{L}_{k-1}^{r_{k-1}^i(u_{k-1})}) \\
  & =  \int_{[0,1]^m} \mathcal{U}(du_{k-1}) \\
  & \quad \quad \cdot \int_{\mathbb{X}} q_k(d x_k \, | \, \mathcal{L}_{k-1}^{r_{k-1}^s(u_{k-1})} \psi_k(\mathcal{L}_{k-1}^{r_{k-1}^s(u_{k-1})}, x_k),
\end{align*}
which give the probabilities in  (\ref{eq:twisted_S}) and line 11 in Algorithm \ref{alg:twisted_pf}.
Next we proceed to draw $U_{k-1}$ conditional on $S_k=s$.
Given $S_k=s$, the distribution for $U_{k-1}$, denoted with $\widetilde{\mathcal{U}}_{k-1}(\cdot \, | \, s)$, is given by
\begin{align*}
& \widetilde{\mathcal{U}}_{k-1}(du_{k-1} \, | \, s) \\
& \propto \int_{\mathbb{X}^n} \mathcal{U}(d u_{k-1}) \psi_k(\mathscr{L}_{k-1}^{r^{s}_{k-1}(u_{k-1})},x_k) q_k(dx_k^s\,|\,\mathscr{L}_{k-1}^{r_{k-1}^s(u_{k-1})}) \\
& \quad  \quad \quad  \cdot \prod_{i \neq s} q_k(x_k^i \, | \, \mathscr{L}_{k-1}^{r_{k-1}^i(u_{k-1})}) \\
& =\mathcal{U}(d u_{k-1}) \int_{\mathbb{X}}  \psi_k(\mathscr{L}_{k-1}^{r^{s}_{k-1}(u_{k-1})},x_k) q_k(dx_k\,|\,\mathscr{L}_{k-1}^{r_{k-1}^s(u_{k-1})}).
\end{align*}
This gives  (\ref{eq:twisted_U}) and line 12 in Algorithm \ref{alg:twisted_pf}.
Finally, given $S_k=s$ and $U_{k-1} = u_{k-1}$, the distribution for $\xi_k$ is proportional to
\begin{align*}
& q_k(d\xi_k^s \, | \, \mathscr{L}_{k-1}^{r_{k-1}^s(u_{k-1})}) \psi_{k}(\mathscr{L}_{k-1}^{r_{k-1}^s(u_{k-1})},\xi_k^s) \\
 & \cdot \prod_{i \neq s} q_k(\xi_k^i \, | \, \mathscr{L}_{k-1}^{r_{k-1}^i(u_{k-1})}).
\end{align*}
This gives $\widetilde{q}_k$ in  (\ref{eq:twisted_q}) and lines 13 to 17 in Algorithm \ref{alg:twisted_pf} for sampling $\xi_k$.

We next show that the expression for $\widetilde{Z}_k$ in Algorithm \ref{alg:twisted_pf} can equivalently be written
\begin{equation}
\widetilde{Z}_k = Z_k \prod_{s=0}^k \phi_s, \quad k \geq 0,
\label{eq:tildeZ}
\end{equation}
where for each $0 \leq s \leq k$, $\phi_s$ is the Radon-Nikodym derivative $d\mathbf{M}_s/d \widetilde{\mathbf{M}}_s$.
The result $\widetilde{\mathbb{E}}[\widetilde{Z}_k] = \mathbb{E}[Z_k]$ then immediately follows from the properties of the Radon-Nikodym derivative. Then, using Proposition \ref{prop:unbiased}, we get  (\ref{eq:twisted_unbiased}).

To compute the Radon-Nikodym derivatives we need to find the normalizing factors in (\ref{eq:particle_filter_t0})-(\ref{eq:particle_filter_tk}).
For $k=0$ the normalization factor is $ \int\psi_0(x) q_0(dx)$ and we get
\begin{align}
\phi_0(\xi_0) & = \frac{d \mathbf{M}_0(\cdot)}{d \widetilde{\mathbf{M}}_0(\cdot)}(\xi_0) = \frac{\int_{\mathbb{X}} \psi_0(x_0)q_0(dx_0)}{\frac{1}{n}\sum_{i=1}^n \psi_0(\xi_0^i)}.\label{eq:phi_0}
\end{align}
For $k>0$, the normalization factor is given by
\begin{align*}
& \int_{[0,1]^m} \int_{\mathbb{X}^n} \frac{1}{n}\sum_{s=1}^n\mathbf{M}_k(d \xi_k, d{u_{k-1}} \, | \, \mathscr{F}_{k-1}) \\
 & \quad \quad \quad  \quad \cdot \psi_{k}(\mathscr{L}_{k-1}^{r_{k-1}^s(u_{k-1})},\xi_k^s) \\
& = \int_{[0,1]^m} \mathcal{U}(du_{k-1}) \\
&\quad  \quad \cdot \frac{1}{n}\sum_{s=1}^n \int_{\mathbb{X}} \psi_k(\mathscr{L}_{k-1}^{r^{s}_{k-1}(u_{k-1})},x_k) q_k(dx_k\,|\,\mathscr{L}_{k-1}^{r_{k-1}^s(u_{k-1})}) \\
& = \mathbb{E}\left[ \frac{1}{n}\sum_{s=1}^n \int_{\mathbb{X}} \psi_k(\mathscr{L}_{k-1}^{r^{s}_{k-1}(U_{k-1})},x_k) \right. \\
& \quad \quad \quad \quad \quad \quad \left. \left. \cdot q_k(dx_k\,|\,\mathscr{L}_{k-1}^{r_{k-1}^s(U_{k-1})})\, \right| \, \mathscr{F}_{k-1}\right] \\
& = \frac{\sum_{i=1}^n W_{k-1}^i \int_{\mathbb{X}} \psi_k(\mathscr{L}_{k-1}^{i},x_k) q_k(dx_k\,|\,\mathscr{L}_{k-1}^{i})}{\sum_{i=1}^n W_{k-1}^i} \\
& = \frac{\sum_{i=1}^n \widetilde{W}_{k-1}^i }{\sum_{i=1}^n W_{k-1}^i}
\end{align*}
where we used Assumption \ref{ass:resampling} and $\widetilde{W}_{k-1}^i$ are given by  (\ref{eq:twisted_w}).
The Radon-Nikodym derivative for $k>0$ is now found to be
\begin{align}
\phi_k(\mathscr{F}_{k-1}, \xi_k) & = \frac{d \mathbf{M}_k(\cdot \, | \, \mathscr{F}_{k-1})}{d \widetilde{\mathbf{M}}_k(\cdot \, | \, \mathscr{F}_{k-1})}(\xi_k) \nonumber  \\
& = \frac{\sum_{i=1}^n \widetilde{W}_{k-1}^i}{\sum_{i=1}^n W_{k-1}^i}\frac{1}{\frac{1}{n} \sum_{i=1}^n \psi_k(\mathscr{L}_k^i)}.\label{eq:phi_k}
\end{align}
Writing out the expression for $\widetilde{Z}_k$ from Algorithm \ref{alg:twisted_pf} and using the expression for $Z_k$ from Algorithm \ref{alg:particle_filter}, we have
\begin{align}
\widetilde{Z}_0 &= \dfrac{\sum_{i=1}^n W_0^i \int_{\mathbb{X}} \psi_0(x_0) q_0(d x_0)}{\sum_{j=1}^n \psi_0(\xi_0^j)}=Z_0 \dfrac{\int_{\mathbb{X}} \psi_0(x_0) q_0(d x_0)}{\frac{1}{n}\sum_{j=1}^n \psi_0(\xi_0^j)},\label{eq:Z_0_derive}
\end{align}
and for $k>0$
\begin{align}
&\widetilde{Z}_k
  =  \widetilde{Z}_{0}  \frac{\sum_{i=1}^n W_k^i}{\sum_{i=1}^n W_{0}^i} \prod_{s=1}^k \frac{\sum_{i=1}^n \widetilde{W}_{s-1}^i}{\sum_{i=1}^n \psi_s(\mathscr{L}_s^i)}\nonumber \\
 &= Z_k \dfrac{\int_{\mathbb{X}} \psi_0(x_0) q_0(d x_0)}{\sum_{j=1}^n \psi_0(\xi_0^j)} \prod_{s=1}^k \frac{\sum_{i=1}^n \widetilde{W}_{s-1}^i}{\sum_{i=1}^n W_{s-1}^i} \frac{1}{\frac{1}{n}\sum_{i=1}^n \psi_s(\mathscr{L}_s^i)},\label{eq:Z_k_derive}
\end{align}
and combining \eqref{eq:Z_0_derive}-\eqref{eq:Z_k_derive} with \eqref{eq:phi_0}-\eqref{eq:phi_k} we observe that \eqref{eq:tildeZ} holds as claimed.

\section{Proof of Theorem \ref{thm:optimal}}
\label{app:proof_theorem2}
With this choice of twisting function, we have the following result for $0 \leq k \leq t-1$
\begin{align*}
\widetilde{W}_{k}^i  & = W_{k}^i \int_{\mathbb{X}} \psi_{k+1}(\mathscr{L}_k^i, x_{k+1})q_{k+1}(x_{k+1} \, | \, \mathscr{L}_k^i)\, dx_{k+1} \\
& = W_{k}^i \int_{\mathbb{X}} f_{k+1}(x_{k+1} \, | \, \xi_{k}^i) p(y_{k+1:T} \, | \, x_{k+1}) \, d x_{k+1} \\
& = W_{k}^i \int_{\mathbb{X}} p(y_{k+1:T}, x_{k+1} \, | \, \xi_{k}^i) \, dx_{k+1} \\
& = W_{k}^i \,p(y_{k+1:t} \, | \, \xi_{k}^i) = \psi_{k}(\mathscr{L}_k^i).
\end{align*}
The final step follows by plugging in $W_k^i$ and noting that $g_k(y_k \, | \, \xi_k^i)p(y_{k+1:t} \, | \, \xi_k^i) = p(y_{k:t} \, | \, \xi_k^i)$.
Furthermore, for $k=t$, we have $\psi_t(\mathscr{L}_t^i) = W_t^i$.

Expanding and rearranging terms in the expression for $\widetilde{Z}_t$ in Algorithm \ref{alg:twisted_pf} we get
\begin{align*}
\widetilde{Z}_{t} & =  \int_{\mathbb{X}} q_0(dx_0)  \psi_{0}(x_{0})\prod_{k=0}^{t-1} \frac{\sum_{i=1}^n W_k^i}{\sum_{i=0}^n \psi_{k}{(\mathscr{L}_k^i)}}\frac{\sum_{i=1}^n \widetilde{W}_k^i}{\sum_{i=1}^n W_k^i} \\
& \quad \quad \cdot \frac{\sum_{i=1}^n W_t}{\sum_{i=1}^n \psi_t(\mathscr{L}_t^i)} \\
 &  = \int_{\mathbb{X}} q_0(dx_0)  \psi_{0}(x_{0})\prod_{k=0}^{t-1} \frac{\sum_{i=1}^n \widetilde{W}_k^i}{\sum_{i=1}^n \psi_{k}{(\mathscr{L}_k^i)}}\frac{\sum_{i=1}^n W_t}{\sum_{i=1}^n \psi_t(\mathscr{L}_t^i)} \\
 & =  \int_{\mathbb{X}} \mu_0(dx_0) p(y_{0:t} \, | \, x_0) = p(y_{0:t}).
\end{align*}

% use section* for acknowledgment
\section*{Acknowledgment}
J. Ala-Luhtala acknowledges financial support from the Tampere University of Technology Doctoral Programme in Engineering and Natural Sciences, Emil Aaltonen foundation and KAUTE foundation. N. Whiteley and K. Heine were partly supported by EPSRC grant EP/K023330/1 and SuSTaIn.

% Can use something like this to put references on a page
% by themselves when using endfloat and the captionsoff option.
%\ifCLASSOPTIONcaptionsoff
%  \newpage
%\fi

% trigger a \newpage just before the given reference
% number - used to balance the columns on the last page
% adjust value as needed - may need to be readjusted if
% the document is modified later
%\IEEEtriggeratref{8}
% The "triggered" command can be changed if desired:
%\IEEEtriggercmd{\enlargethispage{-5in}}

% references section

% can use a bibliography generated by BibTeX as a .bbl file
% BibTeX documentation can be easily obtained at:
% http://www.ctan.org/tex-archive/biblio/bibtex/contrib/doc/
% The IEEEtran BibTeX style support page is at:
% http://www.michaelshell.org/tex/ieeetran/bibtex/
\bibliographystyle{IEEEtran}
% argument is your BibTeX string definitions and bibliography database(s)

%\bibliography{IEEEabrv,bib_twisted}
\bibliography{bib_twisted}

\end{document}